\documentclass[letter]{article}

\usepackage{pdfpages}
\usepackage{fullpage}
\usepackage{textcomp}
\usepackage{algorithm}
\usepackage{algorithmicx}
\usepackage{algpseudocode}
\usepackage{xcolor}
\usepackage{multirow}
\usepackage{amsmath}
\usepackage{amsfonts}
\usepackage{graphicx}
\usepackage{amssymb}

\algtext*{EndWhile}% Remove "end while" text
\algtext*{EndIf}% Remove "end if" text
\algtext*{EndFor}% Remove "end for" text

\algnewcommand{\IIf}[1]{\State\algorithmicif\ #1\ \algorithmicthen}
\algnewcommand{\EndIIf}{\unskip\ \algorithmicend\ \algorithmicif}

\newtheorem{lemma}{Lemma}
\newtheorem{lemma*}[lemma]{Lemma*}
\newtheorem{theorem}[lemma]{Theorem}
\newtheorem{theorem*}[lemma]{Theorem*}
\newtheorem{corollary}[lemma]{Corollary}

\newtheorem{fact}[lemma]{Fact}

\title{\Large Near-Optimal Algorithms for Shortest Paths in Weighted Unit-Disk Graphs\footnote{The work was partially done when Jie Xue was visiting Utah State University. The research of Jie Xue is partially supported by a Doctoral Dissertation Fellowship from the Graduate School of the University of Minnesota.}}

\author{
	Haitao Wang \\ Utah State University \\ \texttt{haitao.wang@usu.edu}
	\and
	Jie Xue \\ University of Minnesota \\ \texttt{xuexx193@umn.edu}
}

\date{}

\begin{document}
	
	\maketitle
	
	\begin{abstract}
		We revisit a classical graph-theoretic problem, the \textit{single-source shortest-path} (SSSP) problem, in weighted unit-disk graphs.
		We first propose an exact (and deterministic) algorithm which solves the problem in $O(n \log^2 n)$ time using linear space, where $n$ is the number of the vertices of the graph.
		This significantly improves the previous deterministic algorithm by Cabello and Jej\v{c}i\v{c} [CGTA'15] which uses $O(n^{1+\delta})$ time and $O(n^{1+\delta})$ space (for any small constant $\delta>0$) and the previous randomized algorithm by Kaplan et al. [SODA'17] which uses $O(n \log^{12+o(1)} n)$ expected time and $O(n \log^3 n)$ space.
		More specifically, we show that if the 2D offline insertion-only (additively-)weighted nearest-neighbor problem with $k$ operations (i.e., insertions and queries) can be solved in $f(k)$ time, then the SSSP problem in weighted unit-disk graphs can be solved in $O(n \log n+f(n))$ time.
		Using the same framework with some new ideas, we also obtain a $(1+\varepsilon)$-approximate algorithm for the problem, using $O(n \log n + n \log^2(1/\varepsilon))$ time and linear space.
		This improves the previous $(1+\varepsilon)$-approximate algorithm by Chan and Skrepetos [SoCG'18] which uses $O((1/\varepsilon)^2 n \log n)$ time and $O((1/\varepsilon)^2 n)$ space.
		More specifically, we show that if the 2D offline insertion-only weighted nearest-neighbor problem with $k_1$ operations in which at most $k_2$ operations are insertions can be solved in $f(k_1,k_2)$ time, then the $(1+\varepsilon)$-approximate SSSP problem in weighted unit-disk graphs can be solved in $O(n \log n+f(n,O(\varepsilon^{-2})))$ time.
		Because of the $\Omega(n \log n)$-time lower bound of the problem (even when approximation is allowed), both of our algorithms are almost optimal.
	\end{abstract}
	
	\section{Introduction}
	Given a set $S$ of $n$ points in the plane, its \textit{unit-disk graph} is an undirected graph in which the vertices are points of $S$ and two vertices are connected by an edge iff the (Euclidean) distance between them is at most 1.
	Unit-disk graphs can be viewed as the intersection graphs of equal-sized disks in the plane, and find many applications such as modeling the topology of ad-hoc communication networks.
	As an important class of geometric intersection graphs, unit-disk graphs have been extensively studied in computational geometry.
	Many problems that are difficult in general graphs have been efficiently solved (exactly or approximately) in unit-disk graphs by exploiting their underlying geometric structures.
	%A \textit{weighted} unit-disk graph is a unit-disk graph where each edge $(a,b)$ has a weight equal to the distance between $a$ and $b$.
	
	In this paper, we consider a classical graph-theoretic problem, the \textit{single-source shortest-path} (SSSP) problem, in unit-disk graphs.
	Given an edge-weighted graph $G = (V,E)$ and a source vertex $s \in V$, the SSSP problem aims to compute shortest paths from $s$ to all other vertices in $G$ (or equivalently a shortest-path tree from $s$).
	In unit-disk graphs, there are two natural ways to weight the edges.
	The first way is to equally weight all the edge (usually called \textit{unweighted} unit-disk graphs), while the second way is to assign each edge $(a,b)$ a weight equal to the (Euclidean) distance between $a$ and $b$ (usually called \textit{weighted} unit-disk graphs).
	The SSSP problem in a general graph has a trivial $\Omega(|E|)$-time lower bound, because specifying the edges of the graph already takes $\Omega(|E|)$ time.
	However, this lower bound does not hold in unit-disk graphs.
	A unit-disk graph (either unweighted or weighted), though having quadratic number of edges in worst case (e.g., all the vertices are very close to each other), can be represented by only giving the locations of its vertices in the plane.
	This linear-complexity representation allows us to solve the SSSP problem without explicitly constructing the graph and hence beat the $\Omega(|E|)$-time lower bound.
	%It is well-known that the SSSP problem (in a general graph) can be solved in $O(|E| + |V| \cdot \log |V|)$ time.
	%However, this algorithm is not efficient when applying to unit-disk graphs, as a unit-disk graph can have quadratic number of edges in worst-case.
	
	In unweighted unit-disk graphs, the SSSP problem is relatively easy, and various algorithms are known for solving it \textit{optimally} in $O(n \log n)$ time 
	%where $n$ is the number of vertices 
	\cite{cabello2015shortest,chan2016all}.
	However, the weighted case 
	%problem in weighted unit-disk graphs 
	is much more challenging.
	Despite of much effort made over years \cite{cabello2015shortest,chan2018approximate,gao2005well,kaplan2017dynamic,roditty2011bounded}, state-of-the-art algorithms are still far away from being optimal.
	%Therefore, the main focus of this paper is on weighted unit-disk graphs.
	In this paper, we present new exact and approximation algorithms for the problem in weighted unit-disk graphs, which significantly improve the previous results and almost match the lower bound of the problem.
	\medskip
	
	\noindent
	\textbf{Organization.}
	The remaining paper is organized as follows.
	In Section~\ref{sec-related}, we discuss the related work and our contributions.
	Section~\ref{sec-notation} presents some notations used throughout the paper.
	%We suggest the reader to read this part carefully before moving on.
	Our exact and approximation algorithms are given in Section~\ref{sec-exact} and~\ref{sec-approx}, respectively.
	%Due to the page limit, some lemma proofs are omitted but can be found in the appendix. 
	
	\subsection{Related Work and Our Contributions} \label{sec-related}
	Besides the SSSP problem, many graph-theoretic problems have also been studied in unit-disk graphs, such as maximum independent set \cite{matsui1998approximation}, maximum clique \cite{clark1990unit}, distance oracle \cite{chan2018approximate,gao2005well}, diameter computing \cite{chan2018approximate,gao2005well}, all-pair shortest paths \cite{chan2016all,chan2017all}, etc.
	Most of these problems have much more efficient solutions in unit-disk graphs than in general graphs.
	
	The SSSP problem in unit-disk graphs has received a considerable attention in the last decades.
	The problem has an $\Omega(n \log n)$-time lower bound even when approximation is allowed, because deciding the connectivity of a unit-disk graph requires $\Omega(n \log n)$ time \cite{cabello2015shortest}.
	In unweighted unit-disk graphs, at least two $O(n \log n)$-time SSSP algorithms were known \cite{cabello2015shortest,chan2016all}, which are optimal.
	If the vertices are pre-sorted by their $x$- and $y$-coordinates, the algorithm in \cite{chan2016all} can solve the problem in $O(n)$ time.
	In weighted unit-disk graphs, the SSSP problem was studied in \cite{cabello2015shortest,chan2018approximate,gao2005well,kaplan2017dynamic,roditty2011bounded}.
	Both exact and approximation algorithms were given to solve the problem in sub-quadratic time.
	For the exact case, the best known results are the deterministic algorithm by Cabello and Jej\v{c}i\v{c} \cite{cabello2015shortest} which uses $O(n^{1+\delta})$ time and $O(n^{1+\delta})$ space (for any small constant $\delta>0$) and the randomized algorithm by Kaplan et al. \cite{kaplan2017dynamic} which uses $O(n \log^{12+o(1)} n)$ expected time and $O(n \log^3 n)$ space.
	For the approximation case, the best known result is the $(1+\varepsilon)$-approximate algorithm by Chan and Skrepetos \cite{chan2018approximate} which uses $O((1/\varepsilon)^2 n \log n)$ time and $O((1/\varepsilon)^2 n)$ space.
	
	In this paper, we first propose an exact SSSP algorithm in weighted unit-disk graphs which uses $O(n \log^2 n)$ time and $O(n)$ space, significantly improving the results in \cite{cabello2015shortest,kaplan2017dynamic}.
	Using the same framework together with some new ideas, we also obtain a $(1+\varepsilon)$-approximate algorithm which uses $O(n \log n + n \log^2 (1/\varepsilon))$ time and $O(n)$ space, improving the result in \cite{chan2018approximate}.
	Table~\ref{tab-results} presents the comparison of our new algorithms with the previous results.
	
	\begin{table}[h]
		\centering
		\begin{tabular}{|c|c|c|c|c|}
			\hline
			\textbf{Type} & \textbf{Source} & \textbf{Time} & \textbf{Space} & \textbf{Rand./Det.} \\
			\hline
			\multirow{4}{*}{Exact} & \cite{roditty2011bounded} & $O(n^{4/3+\delta})$ & $O(n^{1+\delta})$ & Deterministic \\
			\cline{2-5}
			& \cite{cabello2015shortest} & $O(n^{1+\delta})$ & $O(n^{1+\delta})$ & Deterministic \\
			\cline{2-5}
			& \cite{kaplan2017dynamic} & $O(n \log^{12+o(1)} n)$ & $O(n \log^3 n)$ & Randomized \\
			\cline{2-5}
			& {\color{red}Corollary~\ref{cor-exact}} & {\color{red}$O(n \log^2 n)$} & {\color{red}$O(n)$} & {\color{red}Deterministic} \\
			\hline
			\multirow{3}{*}{Approximate} & \cite{gao2005well} & $O((1/\varepsilon)^3 n^{1.5} \sqrt{\log n})$ & $O((1/\varepsilon)^4 n \log n)$ & Deterministic \\
			\cline{2-5}
			& \cite{chan2018approximate} & $O((1/\varepsilon)^2 n \log n)$ & $O((1/\varepsilon)^2 n)$ & Deterministic \\
			\cline{2-5}
			& {\color{red}Corollary~\ref{cor-approximation}} & {\color{red}$O(n \log n+ n \log^2 (1/\varepsilon))$} & {\color{red}$O(n)$} & {\color{red}Deterministic} \\
			\hline
		\end{tabular}
		\caption{Summary of the previous and our new algorithms for SSSP in weighted unit-disk graphs.}
		\label{tab-results}
	\end{table}
	
	More specifically, our algorithms solve the SSSP problem in weighted unit-disk graphs by reducing it to the (2D) offline insertion-only additively-weighted nearest-neighbor (OIWNN) problem, in which we are given a sequence of operations each of which is either an insertion (inserting a weighted point in $\mathbb{R}^2$ to the dataset) or a weighted nearest-neighbor query (asking for the additively-weighted nearest neighbor of a given query point in the dataset) and our goal is to answer all the queries.
	The reductions imply the following results.
	\begin{itemize}
		\item If the OIWNN problem with $k$ operations can be solved in $f(k)$ time, then the exact SSSP problem in weighted unit-disk graphs can be solved in $O(n \log n+f(n))$ time.
		\item If the OIWNN problem with $k_1$ operations in which at most $k_2$ operations are insertions can be solved in $f(k_1,k_2)$ time, then the $(1+\varepsilon)$-approximate SSSP problem in weighted unit-disk graphs can be solved in $O(n \log n+n \log (1/\varepsilon)+f(n,O(\varepsilon^{-2})))$ time.
	\end{itemize}
	Our time bounds in Table~\ref{tab-results} are derived from the above results by arguing that $f(k) = O(k \log^2 k)$ and $f(k_1,k_2) = O(k_1 \log^2 k_2)$.
	Therefore, the bottleneck of our algorithms in fact comes from the OIWNN problem.
	%More efficient algorithms for the OIWNN problem can readily improve the running time of our SSSP algorithms.
	
	As an immediate application, our approximation algorithm can be applied to improve the preprocessing time of the distance oracles in weighted unit-disk graphs given by Chan and Skrepetos~\cite{chan2018approximate}.
	%(see Appendix~\ref{appx-oracle} for the details).
	
	\subsection{Notations} \label{sec-notation}
	In this section, we present the basic notations and concepts used in this paper.
	\medskip
	
	\noindent
	\textbf{Basic notations.}
	Throughout the paper, the notation $\lVert \cdot \rVert$ denotes the Euclidean norm; therefore, for two points $a,b \in \mathbb{R}^2$, $\lVert a-b \rVert$ is the Euclidean distance between $a$ and $b$.
	For a point $a \in \mathbb{R}^2$, we use $\odot_a$ to denote the unit disk (i.e., disk of radius 1) centered at $a$.
	\medskip
	
	\noindent
	\textbf{Graphs.}
	Let $G = (V,E)$ be an edge-weighted undirected graph.
	A path in $G$ is represented as a sequence $\pi = \langle z_1,\dots,z_t \rangle$ where $z_1,\dots,z_t \in V$ and $(z_i,z_{i+1}) \in E$ for all $i \in \{1,\dots,t-1\}$; the \textit{length} of $\pi$ is the sum of the weights of the edges $(z_1,z_2),\dots,(z_{t-1},z_t)$.
	For two vertices $u,v \in V$, we use $\pi_G(u,v)$ to denote the shortest path from $u$ to $v$ in $G$ and use $d_G(u,v)$ to denote the length of $\pi_G(u,v)$.
	We say $v' \in V$ is the $u$-predecessor of $v$ if $(v',v) \in E$ is the last edge of $\pi_G(u,v)$.
	%If two vertices $u,v \in V$ appear adjacently on a path $\pi$ and $u$ is before $v$, we say that $u$ is the \textit{predecessor} of $v$ on $\pi$.
	For two paths $\pi$ and $\pi'$ in $G$ where $\pi$ is from $u$ to $v$ and $\pi'$ is from $v$ to $w$, we denote by $\pi \circ \pi'$ the \textit{concatenation} of $\pi$ and $\pi'$, which is a path from $u$ to $w$ in $G$.
	
	%\subparagraph{Weighted unit-disk graphs.}
	%For a set $S$ of points in $\mathbb{R}^2$, we use $\mathsf{WUDG}(S)$ to denote the weighted unit-disk graph whose vertex set is $S$.
	
	\section{The Exact Algorithm} \label{sec-exact}
	In this section, we describe our exact algorithm. 
	%SSSP algorithm in weighted unit-disk graphs.
	Given a set $S$ of $n$ points in the plane and a source $s \in S$, our goal is to compute a shortest-path tree from $s$ in the weighted unit-disk graph $G$ induced by $S$.
	For all $a \in S$, we use $\lambda_a \in S$ to denote the $s$-predecessor of $a$.
	Specifically, we aim to compute two tables $\text{dist}[\cdot]$ and $\text{pred}[\cdot]$ indexed by the points in $S$, where $\text{dist}[a] = d_G(s,a)$ and $\text{pred}[a] = \lambda_a$.
	
	%Our SSSP algorithm is essentially a variant of the well-known Dijkstra's algorithm.
	%Before discussing our SSSP algorithm, 
	We first briefly review how the well-known Dijkstra's algorithm computes shortest paths from a source $s$ in a graph $G$.
	Initially, the algorithm sets all dist-values to infinity except $\text{dist}[s] = 0$, and sets $A = S$.
	Then it keeps doing the following procedure until $A = \emptyset$.
	
	\begin{enumerate}
		\item Pick the vertex $c \in A$ with the smallest dist-value.
		\item For all $b \in A$ that are neighbors of $c$, update the value $\text{dist}[b]$ using $c$, i.e., $\text{dist}[b] \leftarrow \min\{\text{dist}[b],\text{dist}[c]+w(c,b)\}$, where $w(c,b)$ is the weight of the edge $(c,b)$.
		%If $\text{dist}[b] = \infty$ before refining, add $b$ to $A$.
		\item Remove $c$ from $A$.
	\end{enumerate}
	\medskip
	Directly applying Dijkstra's algorithm to solve the SSSP problem in a weighted unit-disk graph takes quadratic time, since the graph can have $\Omega(n^2)$ edges in worst-case.
	
	Our algorithm will follow the spirit of Dijkstra's algorithm in a high level and exploit many insights of unit-disk graphs in order to achieve a near-linear running time.
	First of all, we (implicitly) build  a grid $\varGamma$ on the plane, which consists of square cells with side-length $1/2$ (a similar grid is also used in~\cite{chan2016all}).
	Assume for convenience that no point in $S$ lies on a grid line, and hence each point in $S$ is contained in exactly one cell of $\varGamma$.
	A \textit{patch} of $\varGamma$ is a square area consisting of $5 \times 5$ cells of $\varGamma$.
	For a point $a \in S$, let $\Box_a$ denote the cell of $\varGamma$ containing $a$ and $\boxplus_a$ denote the patch of $\varGamma$ whose central cell is $\Box_a$.
	For a set $P$ of points in $\mathbb{R}^2$ and a cell $\Box$ (resp., a patch $\boxplus$) of $\varGamma$, define $P_\Box = P \cap \Box$ (resp., $P_\boxplus = P \cap \boxplus$).
	We notice the following simple fact.
	
	\begin{fact}
		For all $a \in S$, we have $S_{\Box_a} \subseteq \mathsf{NB}_G(a) \subseteq S_{\boxplus_a}$, where $\mathsf{NB}_G(a)$ is the set of all neighbors of $a$ in $G$.
	\end{fact}
	
	We compute and store $S_\Box$ (resp., $S_\boxplus$) for all cells $\Box$ (resp., patches $\boxplus$) of $\varGamma$ that contain at least one point in $S$.
	In addition, we associate pointers to each $a \in S$ such that from $a$ one can get access to the stored sets $S_{\Box_a}$ and $S_{\boxplus_a}$.
	The above preprocessing can be easily done in $O(n \log n)$ time and $O(n)$ space after computing $\Box_a$ for all $a \in S$.
	We give in Appendix~\ref{sec:locategrid} a method to compute $\Box_a$ for all $a \in S$ in $O(n \log n)$ time without using the $\mathsf{floor}$ function.
	%After this, we begin our SSSP algorithm.
	
	In order to present our algorithm, we first define a sub-routine \textsc{Update} as follows.
	Suppose we are now at some point of the algorithm.
	If $U$ and $V$ are two subsets of $S$, then the procedure \textsc{Update}$(U,V)$ conceptually does the following.
	\begin{enumerate}
		\item $\text{dist}'[u] \leftarrow \text{dist}[u]$ for all $u \in U$.
		\item $p_v \leftarrow \arg\min_{u \in U \cap \odot_v}\{\text{dist}'[u]+\lVert u-v \rVert\}$ for all $v \in V$.
		\item For all $v \in V$, if $\text{dist}[v] > \text{dist}'[p_v] + \lVert p_v-v \rVert$, then update
		$\text{dist}[v] \leftarrow \text{dist}'[p_v] + \lVert p_v-v \rVert$ and $\text{pred}[v] \leftarrow p_v$.
	\end{enumerate}
	In words, \textsc{Update}$(U,V)$ updates the shortest-path information of the points in $V$ using the shortest-path information of the points in $U$.
	We use lazy update by copying the $\text{dist}[\cdot]$ table to $\text{dist}'[\cdot]$ to guarantee that the order we consider the points in $V$ does not influence the result of the update (note that $U$ and $V$ may not be disjoint).
	However, when $U$ and $V$ are not disjoint, lazy update may result in an inconsistency of shortest-path information, i.e., $\text{dist}[v] > \text{dist}[\text{pred}[v]] + \lVert \text{pred}[v]-v \rVert$ for some $v \in V$ after \textsc{Update}$(U,V)$.
	This can happen when $p_v \in U \cap V$: for example, we update $\text{dist}[v]$ to $\text{dist}'[p_v] + \lVert p_v-v \rVert$ and at the same time $\text{dist}[p_v]$ also gets updated (hence $\text{dist}[p_v] < \text{dist}'[p_v]$), then $\text{dist}[v] > \text{dist}[p_v] + \lVert p_v-v \rVert$ after \textsc{Update}$(U,V)$.
	We call such a phenomenon \textit{data inconsistency}.
	Although \textsc{Update} can result in data inconsistency in general, we shall guarantee it never happens in our algorithm.
	
	The main framework of our algorithm is quite simple, which is presented in Algorithm~\ref{alg-SSSP}.
	%Before discussing the details, we first introduce the basic idea of our algorithm.
	Similarly to Dijkstra's algorithm, we also maintain a subset $A \subseteq S$ during the algorithm and pick the point $c \in A$ with the smallest dist-value in each iteration (line~6).
	The difference is that, instead of using $c$ to update (the shortest-path information of) its neighbors, our algorithm tries to use all points in $A_{\Box_c}$ to update their neighbors (line~8) and then remove them simultaneously from $A$ (line~9).
	%This makes our algorithm different from Dijkstra's.
	However, it is not guaranteed that the shortest-path information of all the points in $A_{\Box_c}$ is correct when $c$ is picked.
	Therefore, before using the points in $A_{\Box_c}$ to update their neighbors, we use an extra procedure to ``correct'' the shortest-path information of these points, which is not needed in Dijkstra's algorithm.
	Surprisingly, we achieve this by simply updating the points in $A_{\Box_c}$ once using the current shortest-path information of their neighbors (line~7).
	
	\begin{algorithm}[tbph]
		\caption{\textsc{SSSP}$(S,s)$}
		\begin{algorithmic}[1]
			\State $\text{dist}[a] \leftarrow \infty$ for all $a \in S$
			\State $\text{pred}[a] \leftarrow \text{NIL}$ for all $a \in S$
			\State $\text{dist}[s] \leftarrow 0$
			\State $A \leftarrow S$
			%\State $I \leftarrow S \backslash \{s\}$
			\While{$A \neq \emptyset$} \Comment{Main loop}
			\State $c \leftarrow \arg \min_{a \in A} \{\text{dist}[a]\}$
			%\State $R \leftarrow S \cap \Box_c$
			\State \textsc{Update}$(A_{\boxplus_c},A_{\Box_c})$ \Comment{First update}
			%\State \textsc{Update}$(r,A)$
			%\State $I \leftarrow I \backslash S_{\Box_c}$
			%\State $A \leftarrow A \cup S_{\Box_c}$
			%
			%\State $Z \leftarrow (A \cup I) \backslash R$
			%\For{$i=1,\dots,|R|$} \Comment{Decomposition loop}
			%\State $b_i \leftarrow \arg \min_{r \in R} \{\text{dist}[r]\}$
			%\State $T_i \leftarrow (Z \backslash \bigcup_{k=1}^{i-1} T_{k}) \cap \odot_{b_i}$
			%\State $R \leftarrow R \backslash \{b_i\}$
			%\EndFor
			%\State $B \leftarrow \emptyset$
			%\For{$i=|R|,\dots,1$} \Comment{Update loop}
			%\State $B \leftarrow B \cup \{b_i\}$
			%\For{$t \in T_i$}
			%\State $p \leftarrow \arg \min_{b \in B} \{\text{dist}[b] + \lVert b-t \rVert\}$
			%\If{$\text{dist}[t] > \text{dist}[p] + \lVert p-t \rVert$}
			%\State $\text{dist}[t] \leftarrow \text{dist}[p] + \lVert p-t \rVert$, $\text{pred}[t] \leftarrow p$
			%\EndIf
			%\EndFor
			%\State \textbf{move} $b_i$ from $A$ to $D$
			%\EndFor
			%\State $R = I \cap \odot_c$
			%\State $B \leftarrow \{b \in S_{\boxplus_c}: \text{dist}[b] = \infty\}$
			\State \textsc{Update}$(A_{\Box_c},A_{\boxplus_c})$ \Comment{Second update}
			%\For{$b \in I \cap S_{\boxplus_c}$}
			%\If{$\text{dist}[b]<\infty$} 
			%\State $I \leftarrow I \backslash \{b\}$
			%\State $A \leftarrow A \cup \{b\}$
			%\EndIf
			%\EndFor
			\State $A \leftarrow A \backslash A_{\Box_c}$
			\EndWhile
			\State \textbf{return} $\text{dist}[\cdot]$ and $\text{pred}[\cdot]$
			%\nolinenumbers
		\end{algorithmic}
		\label{alg-SSSP}
	\end{algorithm}
	
	The correctness of our algorithm is non-obvious.
	Suppose $m$ is the number of the iterations in the main loop.
	Let $c_i$ be the point $c$ picked in the $i$-th iteration.
	%We have the following fact.
	
	\begin{fact} \label{fact-disjoint}
		The points $c_1,\dots,c_m$ belong to different cells in $\varGamma$.
	\end{fact}
	\textit{Proof.}
	Consider two indices $i,j \in \{1,\dots,m\}$ with $i<j$.
	At the moment $c_i$ is chosen (in the $i$-th iteration), $c_j$ must be in $A$.
	However, $c_j \notin A_{\Box_{c_i}}$ for otherwise $c_j$ would have been removed from $A$ (line~9) in the $i$-th iteration.
	Thus, $c_i$ and $c_j$ are in different cells in $\varGamma$.
	\hfill $\Box$
	\medskip
	
	\noindent
	To prove the algorithm correctness, we first show that the dist-values of all points in $S$ are correctly computed eventually.
	Clearly, during the entire algorithm, the dist-values can only decrease and never become smaller than the true shortest-path distances, i.e., we always have $\text{dist}[a] \geq d_G(s,a)$ for all $a \in S$.
	Keeping this in mind, we prove the following lemma.
	\begin{lemma} \label{lem-correct}
		Algorithm~\ref{alg-SSSP} has the following properties. \\
		\textnormal{\bf (1)} When the $i$-th iteration begins, $\textnormal{dist}[a] = d_G(s,a)$ for all $a \in S$ with $d_G(s,a) \leq d_G(s,{c_i})$. \\
		\textnormal{\bf (2)} After the first update of the $i$-th iteration, $\textnormal{dist}[a] = d_G(s,a)$ for all $a \in S_{\Box_{c_i}}$. \\
		\textnormal{\bf (3)} When the $i$-th iteration ends, $\textnormal{dist}[a] = d_G(s,a)$ for all $a \in S$ with $\lambda_a \in S_{\Box_{c_i}}$.
		%$\bullet$ $\text{dist}[a] = d_G(s,a)$ for all $a \in S$ such that $d_G(s,a) \leq d_G(s,c_{i+1})$.
	\end{lemma}
	\textit{Proof.}
	We first notice that the property \textbf{(3)} follows immediately from the property \textbf{(2)} due to the second update.
	Indeed, for a point $a \in S$, if $\lambda_a \in S_{\Box_{c_i}}$, then $a \in S_{\boxplus_{c_i}}$.
	If $a \in A_{\boxplus_{c_i}}$, then the property \textbf{(2)} implies that the second update makes $\text{dist}[a] = d_G(s,a)$.
	If $a \in S_{\boxplus_{c_i}} \backslash A_{\boxplus_{c_i}}$, then $a \in A_{\Box_{c_j}}$ for some $j<i$ (since $a$ got removed from $A$ in a previous iteration) and the property \textbf{(2)} guarantees that $\text{dist}[a] = d_G(s,a)$ after the first update of the $j$-th iteration.
	%If $a \in S_{\Box_{c_i}}$, then $\text{dist}[a] = d_G(s,a)$ even after the correction loop according to the property \textbf{(2)}.
	As such, we only need to verify the first two properties.
	We achieve this using induction on $i$.
	The base case is $i=1$.
	Note that $c_1 = s$ and $d_G(s,{c_1}) = 0$.
	Thus, to see \textbf{(1)}, we only need to guarantee that $\text{dist}[s] = d_G(s,s) = 0$ when the first iteration begins, which is clearly true.
	After the first update of the first iteration, we have $\text{dist}[a] = \lVert s-a \rVert = d_G(s,a)$ for all $a \in S_{\Box_{c_1}}$, hence the property \textbf{(2)} is satisfied.
	Assume the lemma holds for all $i < k$, and we show it also holds in the $k$-th iteration.
	
	To see the property \textbf{(1)}, let $a \in S$ be a point such that $d_G(s,a) \leq d_G(s,{c_k})$.
	Consider the moment when the $k$-th iteration begins.
	Assume for a contradiction that $\text{dist}[a] > d_G(s,a)$ at that time.
	Suppose $\pi_G(s,a) = \langle z_0,z_1,\dots,z_t \rangle$ where $z_0 = s$ and $z_t = a$.
	Define $j$ as the largest index such that $\text{dist}[z_j] = d_G(s,{z_j})$.
	Note that $j \in \{0,\dots,t-1\}$ because $\text{dist}[s] = d_G(s,s)$ and $\text{dist}[a] \neq d_G(s,a)$.
	Therefore, $\text{dist}[z_j] = d_G(s,{z_j}) < d_G(s,a) \leq d_G(s,{c_k}) \leq \text{dist}[c_k]$.
	This implies $z_j \notin A$ (otherwise it contradicts the fact that $c_k$ is the point in $A$ with the smallest dist-value).
	%Also, $z_j \notin I$ as $\text{dist}[z_j] < \infty$.
	It follows $z_j \in S_{\Box_{c_i}}$ for some $i<k$, as it got removed from $A$ in some previous iteration.
	Then by our induction hypothesis and the property \textbf{(3)}, we have $\text{dist}[z_{j+1}] = d_G(s,z_{j+1})$ at the end of the $i$-th iteration and thus at the beginning of the $k$-th iteration, because $\lambda_{z_{j+1}} = z_j$.
	However, this contradicts the fact that $\text{dist}[z_{j+1}] > d_G(s,z_{j+1})$.
	As such, $\text{dist}[a] = d_G(s,a)$ when the $k$-th iteration begins.
	
	\begin{figure}[t]
		\begin{minipage}[t]{\linewidth}
			\begin{center}
				\includegraphics[totalheight=1.3in]{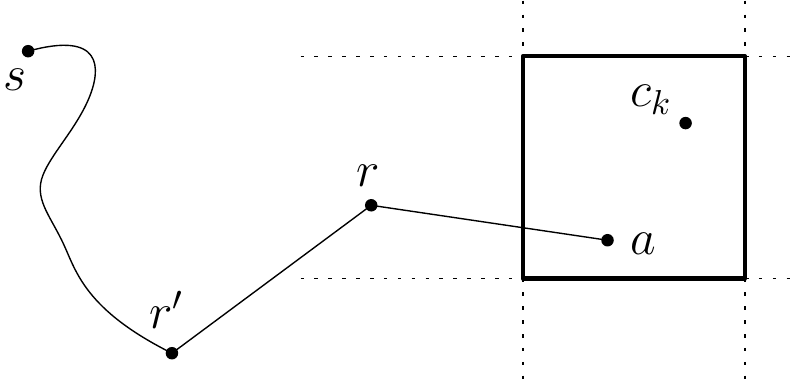}
				\caption{\footnotesize
					Illustrating points $a$, $r$, and $r'$. The solid path is $\pi_G(s,a)$. The solid square is $\Box_{c_k}$. }
				\label{fig:predecessor}
			\end{center}
		\end{minipage}
	\end{figure}
	
	Next, we prove the property \textbf{(2)}.
	For convenience, in what follows, we use $A$ to denote the set $A$ during the $k$-th iteration (before line~9).
	We have $S_{\Box_{c_k}} = A_{\Box_{c_k}}$, since $A = S \backslash (\bigcup_{i=1}^{k-1} S_{\Box_{c_i}})$ and $c_k \notin \Box_{c_i}$ for all $i<k$ by Fact~\ref{fact-disjoint}.
	Let $a \in A_{\Box_{c_k}}$ be a point and $r = \lambda_a$.
	We want to show that $\text{dist}[a] = d_G(s,a)$ after the first update of the $k$-th iteration.
	If $r \notin A$, then $r$ got removed from $A$ in the $i$-th iteration for some $i<k$, namely, $r \in S_{\Box_{c_i}}$.
	By our induction hypothesis and the property \textbf{(3)}, we have $\text{dist}[a] = d_G(s,a)$ at the end of $i$-th iteration (and thus in all the next iterations).
	So assume $r \in A$ (this implies that $r\neq s$ and thus $\lambda_r$ exists).
	In this case, a key observation is that before the first update of the $k$-th iteration, $\text{dist}[r] = d_G(s,r)$.
	To see this, let $r' = \lambda_r$ (e.g., see Fig.~\ref{fig:predecessor}).
	%If $r = s$, then $\text{dist}[r] = d_G(s,r) = 0$.
	%If $r \neq s$, let $r'$ be the $s$-predecessor of $r$.
	Note that $\lVert r' - a \rVert > 1$, otherwise the path $\pi_G(s,r') \circ \langle r',a \rangle$ would be shorter than $\pi_G(s,r') \circ \langle r',r,a \rangle = \pi_G(s,a)$, contradicting the fact that $\pi_G(s,a)$ is the shortest path from $s$ to $a$.
	It follows that
	\begin{equation*}
	d_G(s,a) = d_G(s,r') + d_G(r',a) \geq d_G(s,r') + \lVert r' - a \rVert > d_G(s,r') + 1.
	\end{equation*}
	On the other hand, since $a \in \Box_{c_k}$, we have
	\begin{equation*}
	d_G(s,a) \leq d_G(s,c_k) + d_G(c_k,a) = d_G(s,c_k) + \lVert c_k - a \rVert \leq d_G(s,c_k) + 1.
	\end{equation*}
	Therefore, $d_G(s,r') < d_G(s,c_k)$, and by the property \textbf{(1)} we have $\text{dist}[r'] = d_G(s,r')$ when the $k$-th iteration begins.
	This further implies $r' \notin A$, since $\text{dist}[r'] = d_G(s,r') < d_G(s,c_k) = \text{dist}[c_k]$ when the $k$-th iteration begins.
	Hence, $r'$ got removed from $A$ in the $i$-th iteration for some $i<k$.
	Using our induction hypothesis and the property \textbf{(3)}, we have $\text{dist}[r] = d_G(s,r)$ at the end of the $i$-th iteration (and thus in all the next iterations).
	%Now we see that before the first update of the $k$-th iteration, $\text{dist}[r] = d_G(s,r)$ for all $r \in S$ that is the $s$-predecessor of some point in $S_{\Box_{c_k}}$.
	Note that $r \in \boxplus_{c_k}$, because $r \in \odot_a$.
	We further have $r \in A_{\boxplus_{c_k}}$, as we assumed $r \in A$.
	Hence, the first update of the $k$-th iteration makes $\text{dist}[a] = d_G(s,a)$.
	%If $r \in S_{\boxplus_{c_k}} \backslash A_{\boxplus_{c_k}}$, then $r \in A_{\Box_{c_i}}$ for some $i<k$ because it got removed from $A$ in some previous iteration.
	%By our induction hypothesis and the property \textbf{(3)}, we have $\text{dist}[a] = d_G(s,a)$ at the end of the $i$-th iteration and thus in all the next iterations.
	%As such, after the first update, the dist-values of all $a \in S_{\Box_{c_k}}$ become correct.
	This proves the property \textbf{(2)}. 
	%and also the entire lemma.
	\hfill $\Box$
	\medskip
	
	%The correctness of Algorithm~\ref{alg-SSSP} straightforwardly follows from the above lemma.
	Lemma~\ref{lem-correct} implies that $\text{dist}[a] = d_G(s,a)$ for all $a \in S$ at the end of Algorithm~\ref{alg-SSSP}.
	Indeed, any point $a \in S$ belongs to $S_{\Box_{c_i}}$ for some $i \in \{1,\dots,m\}$, thus the property \textbf{(2)} of Lemma~\ref{lem-correct} guarantees $\text{dist}[a] = d_G(s,a)$.
	%We show $\text{dist}[a] = d_G(s,a)$ at the end of Algorithm~\ref{alg-SSSP}.
	%Suppose $\pi_G(s,a) = \langle z_0,z_1,\dots,z_t \rangle$ where $z_0 = s$ and $z_t = a$.
	%Let $j$ be the largest index such that $\text{dist}[z_j] = d_G(s,z_j)$ at the end of the algorithm.
	%Since $\text{dist}[z_j] < \infty$, we know that $z_j$ got removed from $A$ in some iteration.
	%In other words, $z_j \in S_{\Box_{c_i}}$ for some $i \in \{1,\dots,m\}$.
	%If $j<t$, by the property \textbf{(3)} of Lemma~\ref{lem-correct}, $\text{dist}[z_{j+1}] = d_G(s,z_{j+1})$ at the end of the $i$-th iteration, which contradicts the definition of $j$.
	%Therefore, we have $j=t$ and $\text{dist}[a] = d_G(s,a)$ at the end of Algorithm~\ref{alg-SSSP}.
	Next, we check the correctness of the $\text{pred}[\cdot]$ table.
	We want $\text{dist}[a] = \text{dist}[\text{pred}[a]] + \lVert \text{pred}[a]-a \rVert$ for all $a \in S$.
	However, as mentioned before, the sub-routine \textsc{Update} in general may result in data inconsistency, making this equation false.
	The next lemma shows this can not happen in our algorithm. 
	\begin{lemma}\label{dataconsistAlgo1}
		At any moment of Algorithm~\ref{alg-SSSP}, we always have $\textnormal{dist}[a] = \textnormal{dist}[\textnormal{pred}[a]] + \lVert \textnormal{pred}[a]-a \rVert$ for all $a \in S$.
	\end{lemma}
	\textit{Proof.}
	First, we notice that at any moment of Algorithm~\ref{alg-SSSP}, $\text{dist}[a] \geq \text{dist}[\text{pred}[a]] + \lVert \text{pred}[a]-a \rVert$ for all $a \in S$.
	Indeed, after the procedure \textsc{Update}$(U,V)$, the only data inconsistency that can happen is $\text{dist}[v] > \text{dist}[\text{pred}[v]] + \lVert \text{pred}[v]-v \rVert$ for some $v \in V$.
	So it suffices to show that $\text{dist}[a] \leq \text{dist}[\text{pred}[a]] + \lVert \text{pred}[a]-a \rVert$ for all $a \in S$ at any moment of Algorithm~\ref{alg-SSSP}.
	In fact, we only need to check this after the two update steps, since the $\text{dist}[\cdot]$ and $\text{pred}[\cdot]$ tables only change in the two update steps.
	After each first update, for all $a \in A_{\Box_c}$, we have $\textnormal{dist}[a] = d_G(s,a)$ by the property \textbf{(2)} of Lemma~\ref{lem-correct} and thus $\textnormal{dist}[a] \leq \text{dist}[\text{pred}[a]] + \lVert \text{pred}[a]-a \rVert$ because $\text{dist}[\text{pred}[a]] \geq d_G(s,\text{pred}[a])$ and $d_G(s,a) \leq d_G(s,\text{pred}[a]) + \lVert \text{pred}[a]-a \rVert$.
	Therefore, no data inconsistency happens in the first update.
	In each second update, only the shortest-path information of the points in $A_{\boxplus_c} \backslash A_{\Box_c}$ can be updated (because the dist-values of the points in $A_{\Box_c}$ are already correct after the first update).
	This says the second update is equivalent to \textsc{Update}$(A_{\Box_c},A_{\boxplus_c} \backslash A_{\Box_c})$.
	Since $A_{\Box_c}$ and $A_{\boxplus_c} \backslash A_{\Box_c}$ are disjoint, the second update cannot result in data inconsistency.
	In sum, no data inconsistency occurs during Algorithm~\ref{alg-SSSP}, i.e., we always have $\text{dist}[a] = \text{dist}[\text{pred}[a]] + \lVert \text{pred}[a]-a \rVert$ for all $a \in S$.
	\hfill $\Box$
	\medskip
	
	Now we see that Algorithm~\ref{alg-SSSP} correctly computes shortest paths from $s$.
	However, it is still not clear why simultaneously processing all points in one cell in each iteration makes our algorithm faster than the standard Dijkstra's algorithm.
	In what follows, we focus on the time complexity of the algorithm.
	At this point, let us ignore the two \textsc{Update} sub-routines and show how to efficiently implement the remaining part of the algorithm.
	In each iteration, all the work can be done in constant time except lines~6 and 9.
	To efficiently implement lines~6 and 9, we maintain the set $A$ in a (balanced) binary search tree using the dist-values as keys.
	In this way, line~6 can be done in $O(\log n)$ time, and lines~9 can be done in $O(|S_{\Box_c}| \cdot \log n)$ time.
	%Line~20 takes $O(\log n)$ time for each $b \in I \cap S_{\boxplus_c}$, so the total time is $O(|S_{\boxplus_c}| \cdot \log n)$.
	Note that whenever the dist-value of a point in $A$ is updated, we also need to update the binary search tree in $O(\log n)$ time.
	This occurs in the two \textsc{Update} sub-routines, which has at most $O(|S_{\Box_c}|+|S_{\boxplus_c}|) = O(|S_{\boxplus_c}|)$ modifications of the dist-values.
	Therefore, the time for updating the binary search tree is $O(|S_{\boxplus_c}| \cdot \log n)$.
	%To maintain the set $I$, we do not need any data structure.
	%Instead, we assign each point in $S$ a mark to indicate if it is contained in $I$.
	%In this way, to remove a point from $I$, we only need to modify its mark.
	%Therefore, line~14 can be done in $O(|S_{\Box_c}|)$ time.
	%Line~20 takes $O(1)$ time for each $b \in I \cap S_{\boxplus_c}$ and $O(|S_{\boxplus_c}|)$ time in total.
	%To do the for-loop in line~17-20, we need to find the points in $I \cap S_{\boxplus_c}$, which can be achieved by scanning all points in $S_{\boxplus_c}$ taking $O(|S_{\boxplus_c}|)$ time.
	To summarize, the time cost of the $i$-th iteration, without the \textsc{Update} sub-routines, is $O(|S_{\boxplus_{c_i}}| \cdot \log n)$.
	Since $\sum_{i=1}^m |S_{\boxplus_{c_i}}| \leq 25 n$ by Fact~\ref{fact-disjoint}, the overall time is $O(n \log n)$.
	In the following two sections, we shall consider the time complexities of the two \textsc{Update} sub-routines. 
	To efficiently implement the first \textsc{Update} is relatively easy, while the second one is more challenging.
	
	\subsection{First Update} \label{sec-correction}
	In this section, we show how to implement the first update (line~7) in $O(|S_{\boxplus_c}| \cdot \log n)$ time.
	As mentioned before, we can obtain the points in $S_{\boxplus_c}$ using the pointer associated to $c$, and then further find the points in $A_{\boxplus_c}$ and $A_{\Box_c}$.
	After this, we do $\text{dist}'[a] \leftarrow \text{dist}[a]$ for all $a \in A_{\boxplus_c}$.
	To implement \textsc{Update}$(A_{\boxplus_c},A_{\Box_c})$, the critical step is to find, for every $r \in A_{\Box_c}$, a point $p \in A_{\boxplus_c} \cap \odot_r$ that minimizes $\text{dist}'[p]+\lVert p-r \rVert$.
	This is equivalent to searching the weighted nearest-neighbor of $r$ in the unit disk $\odot_r$ (if we regard $A_{\boxplus_c}$ as a weighted dataset where the weight of each point equals its dist$'$-value).
	Unfortunately, it is currently not known how to efficiently solve this problem.
	%Here, we regard line~8 as part of the correction loop.
	%Explicitly implementing line~8 is time-consuming as it copies the entire $\text{dist}[\cdot]$ table to $\text{dist}'[\cdot]$.
	%However, it is easy to see that we do not need to explicitly copy the table.
	%We shall briefly argue this at the end of the section.
	%At this point, let us assume the $\text{dist}'[\cdot]$ table is already in hand and consider how to implement the correction loop.
	Therefore, we need to exploit some special property of the problem in hand.
	An observation here is that $c$ is the point in $A_{\boxplus_c}$ with the smallest dist$'$-value and all the points in $A_{\Box_c}$ are of distance at most 1 to $c$ (because $c \in A_{\Box_c}$).
	Using this observation, we prove the following key lemma.
	\begin{lemma}\label{lem:firstUpdate}
		Before the first update of each iteration, for all $r \in A_{\Box_c}$, we have 
		\begin{equation*}
		\arg \min_{a \in A_{\boxplus_c} \cap \odot_r} \{\textnormal{dist}'[a] + \lVert a-r \rVert\} = \arg \min_{a \in A_{\boxplus_c}} \{\textnormal{dist}'[a] + \lVert a-r \rVert\}.
		\end{equation*}
	\end{lemma}
	\textit{Proof.}
	\begin{figure}[t]
		\begin{minipage}[t]{\linewidth}
			\begin{center}
				\includegraphics[totalheight=1.8in]{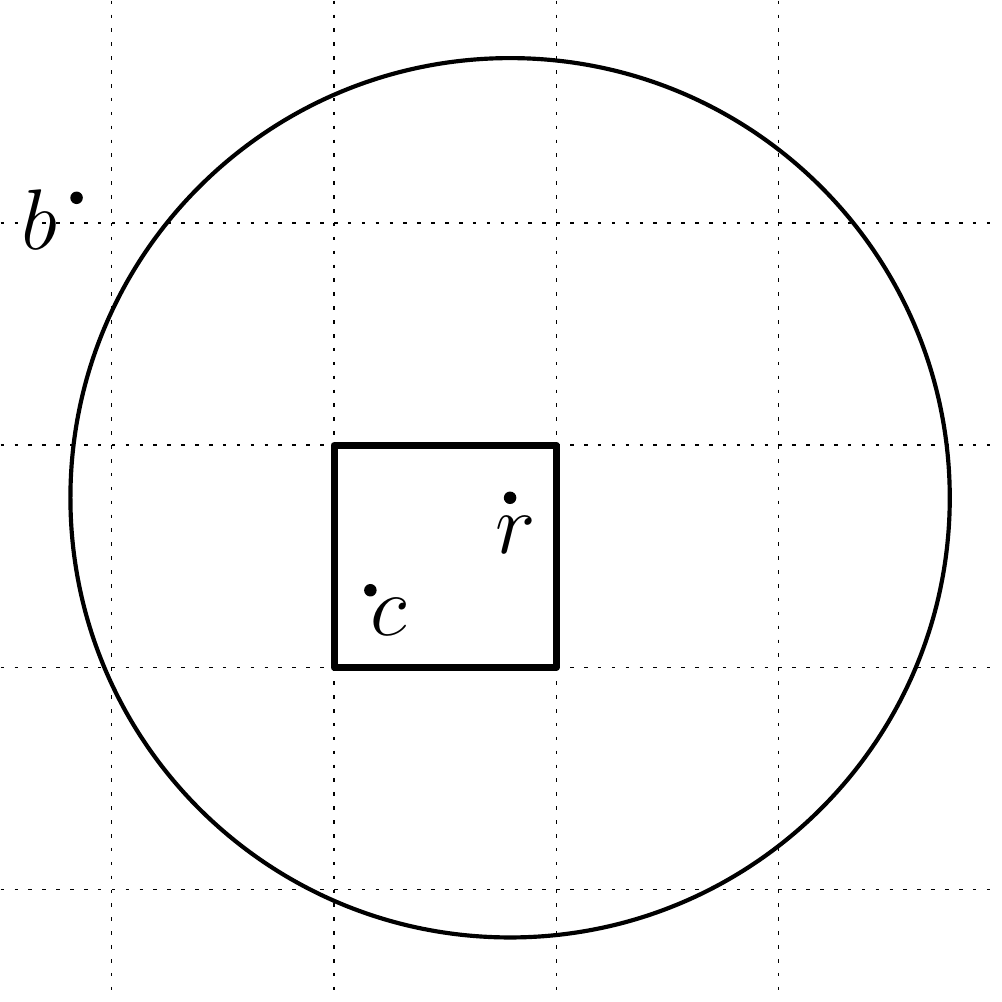}
				\caption{\footnotesize
					Illustrating the proof of Lemma~\ref{lem:firstUpdate}. The solid square is $\Box_c$ and the solid circle is $\odot_r$.}
				%The point $b$ is outside $\odot_r$.}
				\label{fig:firstUpdate}
			\end{center}
		\end{minipage}
	\end{figure}
	Let $p = \arg \min_{a \in A_{\boxplus_c} \cap \odot_r} \{\text{dist}'[a] + \lVert a-r \rVert\}$.
	%Note that $A_{\boxplus_c} \cap \odot_r \subseteq A_{\boxplus_c}$.
	Define $B = A_{\boxplus_c} \backslash (A_{\boxplus_c} \cap \odot_r)$.
	It suffices to show that $\text{dist}'[p] + \lVert p-r \rVert < \text{dist}'[b] + \lVert b-r \rVert$ for all $b \in B$.
	Fix a point $b \in B$ (e.g., see Fig.~\ref{fig:firstUpdate}).
	We have $\lVert b-r \rVert > 1$ by construction.
	On the other hand, since $r \in A_{\Box_c}$, we have $c \in \odot_r$ and hence $\lVert c-r \rVert \leq 1$.
	Furthermore, $\text{dist}'[c] \leq \text{dist}'[b]$, because $b \in A$ and $c$ is the point in $A$ with the smallest dist-value (as well as the smallest dist$'$-value).
	It follows that
	\begin{equation*}
	\text{dist}'[p] + \lVert p-r \rVert \leq \text{dist}'[c] + \lVert c-r \rVert < \text{dist}'[b] + \lVert b-r \rVert,
	\end{equation*}
	where the first ``$\leq$'' follows from the definition of $p$ and the fact that $c \in A_{\Box_c} \cap \odot_r$.
	\hfill $\Box$
	\medskip
	
	\noindent
	The above lemma makes the problem easy.
	Indeed, for every $r \in A_{\Box_c}$, we only need to find a point $p \in A_{\boxplus_c}$ that minimizes $\text{dist}'[p]+\lVert p-r \rVert$ and Lemma~\ref{lem:firstUpdate} guarantees that $p \in \odot_r$.
	This is just the standard (additively-)weighted nearest-neighbor search, which can be solved by building a weighted Voronoi Diagram (WVD) on $A_{\boxplus_c}$ and then querying for each $r \in A_{\Box_c}$.
	Building the WVD takes $O(|A_{\boxplus_c}| \cdot \log |A_{\boxplus_c}|)$ time and linear space~\cite{ref:FortuneA87}, and each query can be answered in $O(\log |A_{\boxplus_c}|)$ time.
	The last step, updating the dist-values and pred-values of the points in $A_{\Box_c}$, is easy.
	So the first update of the $i$-th iteration can be done in $O(|S_{\boxplus_{c_i}}| \cdot \log n)$ time.
	Since $\sum_{i=1}^{m} |S_{\boxplus_{c_i}}| \leq 25n$, the total time for the first update is $O(n \log n)$.
	%Here we also notice that in our implementation line~8 is in fact not needed.
	%The $\text{dist}'[\cdot]$ table is only used to keep the original dist-values of points, as the $\text{dist}[\cdot]$ table may change during the correction loop.
	%Alternatively, we can just use the original dist-values of the points in $A_{\boxplus_c}$ as weights to build the WVD.
	%Note that although we change the $\text{dist}[\cdot]$ table during the correction loop, the weights in the WVD are never modified.
	
	\subsection{Second Update} \label{sec-refine}
	In this section, we consider the second update (line~8) in Algorithm~\ref{alg-SSSP}.
	Unfortunately, the trick used in the first update does not apply, which makes the second update more difficult.
	Here we design a more general algorithm, which can implement \textsc{Update}$(U,V)$ for arbitrary subsets $U,V \subseteq S$ in $O(f(k)+ k \log k)$ time where $k = |U|+|V|$ and $f(k)$ is the time cost of the OIWNN problem with $k$ operations (i.e., insertions and queries).
	%Let $U$ and $V$ be two disjoint subsets of $S$ where the dist-values of the points in $U$ are already computed.
	%Recall that the procedure $\textsc{Refine}(U,V)$ updates the dist-values of all points $v \in V$ as
	%\begin{equation*}
	%    \begin{array}{c}
	%        \text{dist}[v] \leftarrow \min\{\text{dist}[v],\min_{u \in U \cap \mathsf{NB}_G(v)}\{\text{dist}[u]+\lVert u-v \rVert\}\}
	%    \end{array}
	%\end{equation*}
	%and also updates $\text{pred}[v]$ accordingly.
	%Note that $U \cap \mathsf{NB}_G(v) = U \cap \odot_v$, since $G$ is the weighted unit-disk graph.
	%The straightforward implementation of $\textsc{Refine}(U,V)$ requires to compute $\text{dist}[u]+\lVert u-v \rVert$ for all pairs $(u,v)$ where $u \in U$ and $v \in V$, which takes $O(|U| \cdot |V|)$ time.
	%This is unsatisfying as it can result in a quadratic running time of Algorithm~\ref{alg-SSSP}.
	%In what follows, we describe an algorithm to implement $\textsc{Refine}(U,V)$ in $O(f(k)+ k \log k)$ time where $k = |U|+|V|$ and $f(k)$ is the time cost of the offline insertion-only weighted nearest-neighbor problem for $k$ operations.
	The framework of the algorithm is presented in Algorithm~\ref{alg-refine}.
	After copying $\text{dist}[\cdot]$ to $\text{dist}'[\cdot]$, we first sort the points in $U$ in increasing order of their dist$'$-values (line~2).
	Then we compute $|U|$ disjoint subsets $V_1,\dots,V_{|U|}$ of $V$ (line~4), where $V_i$ consists of the points contained in $\odot_{u_i}$ but not contained in $\odot_{u_j}$ for any $j<i$.
	Note that $\bigcup_{i=1}^{|U|} V_i$ consists of all the points in $V$ who have neighbors in $U$, and hence we only need to update the shortest-path information of these points.
	For each point $v \in V_i$, what we do is to find its weighted nearest-neighbor $p$ in $\{u_i,\dots,u_{|U|}\}$ where the weights are the dist-values (line~9), and update the shortest-path information of $v$ by attempting to use $p$ as predecessor (line~10-12).
	
	\begin{algorithm}[tbph]
		\caption{\textsc{Update}$(U,V)$}
		\begin{algorithmic}[1]
			\State $\text{dist}'[u] \leftarrow \text{dist}[u]$ for all $u \in U$.
			\State Sort the points in $U = \{u_1,\dots,u_{|U|}\}$ such that $\text{dist}'[u_1] \leq \cdots \leq \text{dist}'[u_{|U|}]$
			\For{$i=1,\dots,|U|$}
			%\State $u_i \leftarrow \arg \min_{u \in U'} \{\text{dist}[u]\}$
			\State $V_i \leftarrow \{v \in V: v \in \odot_{u_i} \text{ and } v \notin \odot_{u_j} \text{ for all } j<i\}$
			%\State $V \leftarrow V \backslash V_i$
			%\State $U' \leftarrow U' \backslash \{u_i\}$
			\EndFor
			\State  $B \leftarrow \emptyset$
			\For{$i=|U|,\dots,1$}
			\State $B \leftarrow B \cup \{u_i\}$
			\For{$v \in V_i$}
			\State $p \leftarrow \arg \min_{b \in B} \{\text{dist}'[b] + \lVert b-v \rVert\}$
			\If{$\text{dist}[v] > \text{dist}'[p] + \lVert p-v \rVert$}
			\State $\text{dist}[v] \leftarrow \text{dist}'[p] + \lVert p-v \rVert$
			\State $\text{pred}[v] \leftarrow p$
			%\nolinenumbers
			\EndIf
			\EndFor
			\EndFor
		\end{algorithmic}
		\label{alg-refine}
	\end{algorithm}
	
	We first prove the correctness of Algorithm~\ref{alg-refine}.
	Consider a point $v \in V_i$.
	The purpose of \textsc{Update}$(U,V)$ is to find the weighted nearest-neighbor of $v$ in $U \cap \odot_v$ (and use it to update the shortest-path information of $v$), while what we find in line~9 is the weighted nearest-neighbor $p$ in $\{u_i,\dots,u_{|U|}\}$.
	We notice that $U \cap \odot_v \subseteq \{u_i,\dots,u_{|U|}\}$ because $v \notin \odot_{u_j}$ for all $j < i$ by the definition of $V_i$.
	Therefore, we only need to show that the point $p$ computed by line~9 is contained in $U \cap \odot_v$.
	
	\begin{lemma}
		At line~9 of Algorithm~\ref{alg-refine}, we have $p \in U \cap \odot_v$.
	\end{lemma}
	\textit{Proof.}
	Clearly, we have $p \in U$ since $B = \{u_i,\dots,u_{|U|}\} \subseteq U$.
	It suffices to show $\lVert p-v \rVert \leq 1$.
	Assume for a contradiction that $\lVert p-v \rVert > 1$.
	We have $\lVert u_i-v \rVert \leq 1$ since $v \in V_i$.
	Furthermore, $\text{dist}'[u_i] \leq \text{dist}'[p]$ because $p \in \{u_i,\dots,u_{|U|}\}$ and $\text{dist}'[u_i] \leq \text{dist}'[u_j]$ for all $j \geq i$.
	Hence,
	\begin{equation*}
	\text{dist}'[u_i] + \lVert u_i-v \rVert \leq \text{dist}'[u_i]+1 \leq \text{dist}'[p]+1 < \text{dist}'[p] + \lVert p-v \rVert,
	\end{equation*}
	which contradicts the fact that $p$ is the weighted nearest-neighbor of $v$ in $\{u_i,\dots,u_{|U|}\}$.
	\hfill $\Box$
	\medskip
	
	Next, we analyze the time complexity of Algorithm~\ref{alg-refine}.
	At the beginning, we need to sort the points in $U$ in increasing order of their dist-values, which can be done in $O(|U| \cdot \log |U|)$ time and hence $O(k \log k)$ time.
	%The loop in line 2-4 requires to do unit-disk range reporting (line~3) with deletions (line~4).
	%We shall show later how to implement this loop efficiently.
	Algorithm~\ref{alg-refine} basically consists of two loops.
	We first consider the second loop (line 6-12).
	In this loop, what we do is weighted nearest-neighbor search on $B$ (line~9) with insertions (line~7), where the weight of each point $b \in B$ is $\text{dist}'[b]$.
	Note that all insertions and queries here are offline, since the points $u_1,\dots,u_{|U|}$ and the sets $V_1,\dots,V_{|U|}$ are already known before the loop.
	We have $|U|$ insertions and $|V|$ queries, and hence $k$ operations in total.
	Recall that $f(k)$ is the time for solving the OIWNN problem with $k$ operations.
	So this loop takes $f(k)$ time.
	
	Now we consider the first loop (line 3-4).
	This loop requires us to compute $V_i$, the subset of $V$ consisting of the points contained in $\odot_{u_i}$ but not contained in $\odot_j$ for all $j<i$, for $i \in \{1,\dots,|U|\}$.
	We have the following lemma. 
	With the lemma, \textsc{Update}$(U,V)$ can be done in $O(f(k)+k \log k)$ time.
	
	\begin{lemma}\label{lem:firstLoopAlgo1}
		The first loop (line 3-4) of Algorithm~\ref{alg-refine} takes $O(k \log k)$ time where $k = |U|+|V|$.
	\end{lemma}
	
	\subsubsection{Proof of Lemma~\ref{lem:firstLoopAlgo1}}
	
	%In the following, we show that this can be done in $O(k\log k)$ time and $O(k)$ space.
	We prove Lemma~\ref{lem:firstLoopAlgo1} in this section.
	To compute $V_1,\dots,V_{|U|}$, it suffices to compute for each point $v \in V$ the smallest index $i(v)$ such that $\odot_{u_{i(v)}}$ contains $v$, since $V_i = \{v \in V: i(v) = i\}$.
	To this end, we first consider an easy case in which all the points in $U$ are contained in one cell $\Box$ in $\varGamma$ (in fact, this is already sufficient for our algorithm because we have $U = S_{\Box_c}$ in the second $\textsc{Update}$ sub-routine used in Algorithm~\ref{alg-SSSP}).
	Define the top/bottom/left/right bounding line of $\Box$ as the line that contains the top/bottom/left/right boundary of $\Box$, respectively.
	For the points $v \in V$ such that $v \in \Box$, we always have $i(v) = 1$ because $\Box \subseteq \odot_{u_1}$.
	Thus, we only need to consider the points in $V$ that are outside $\Box$.
	A point outside $\Box$ can be separated from $\Box$ by one of the four bounding lines of $\Box$.
	So the problem is reduced to computing $i(v)$ for all $v \in V$ above the top bounding line $l$ of $\Box$.
	To this end, we create a subdivision $\varPhi$ of the halfplane $H$ above $l$ as follows.

	\begin{figure}[t]
		\begin{minipage}[t]{\linewidth}
			\begin{center}
				\includegraphics[totalheight=0.9in]{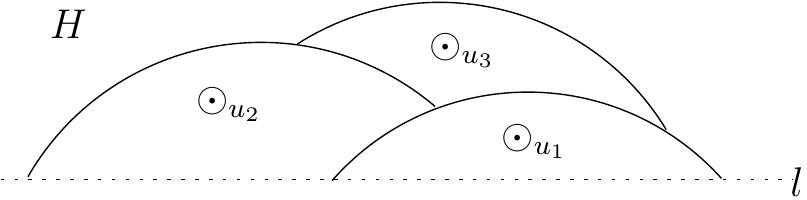}
				\caption{\footnotesize
					Illustrating the subdivision obtained by overlaying three unit-disks $\odot_{u_1}$, $\odot_{u_2}$, and $\odot_{u_3}$.}
				\label{fig:subdivision}
			\end{center}
		\end{minipage}
	\end{figure}

	%Let $\phi_i$ be the part of $\odot_{u_i}$ above $l$.
	Roughly speaking, $\varPhi$ is obtained by overlaying $\odot_{u_1},\dots,\odot_{u_{|U|}}$ in order (e.g., see Fig.~\ref{fig:subdivision}).
	Formally, we begin with a trivial subdivision $\varPhi_0$ of $H$, which consists of only one face, the entire $H$.
	We denote this face by $F_0$ and call it the \textit{outer face} of $\varPhi_0$.
	Suppose now the subdivision $\varPhi_{i-1}$ is defined, which has an outer face $F_{i-1}$ equal to the complement of $H \cap (\bigcup_{j=1}^{i-1} \odot_{u_j})$ in $H$ (note that $F_{i-1}$ is connected).
	We then construct a new subdivision $\varPhi_i$ from $\varPhi_{i-1}$ by decomposing $F_{i-1}$ using $\odot_{u_i}$.
	Specifically, $F_{i-1}$ is decomposed into several new faces in $\varPhi_i$, one of which is $F_{i-1} \backslash \odot_{u_i}$ and the others are the connected components of $F_{i-1} \cap \odot_{u_i}$.
	The face $F_{i-1} \backslash \odot_{u_i}$, which is the complement of $H \cap (\bigcup_{j=1}^i \odot_{u_j})$ in $H$, becomes the outer face $F_i$ of $\varPhi_i$.
	We assign a label $i$ to those new faces corresponding to the connected components of $F_{i-1} \cap \odot_{u_i}$.
	In this way, we obtain a sequence $\varPhi_0,\dots,\varPhi_{|U|}$ of subdivisions of $H$ and define $\varPhi = \varPhi_{|U|}$.
	One can easily verify that, for a point $v \in V \cap H$, if the face of $\varPhi$ containing $v$ is labeled as $i$, then $i(v) = i$.
	Therefore, computing $i(v)$ can be done by a point location in $\varPhi$.
	In what follows, we study the complexity $|\varPhi|$ of $\varPhi$ and how to construct $\varPhi$ efficiently.
	To this end, we need a basic geometric observation.
	Consider a set $A$ of points in $\Box$.
	Let $\xi(A)$ be the boundary of (the closure of) $H \backslash \bigcup_{a \in A} \odot_a$.
	It is clear that $\xi(A)$ is an $x$-monotone curve consisting of ``pieces'', where the leftmost/rightmost pieces are horizontal rays and each of the other pieces is a portion of the boundary of $\odot_a$ for some $a \in A$ (we say the piece is \textit{contributed} by $a$ in this case).
	We make the following observation.
	\begin{fact} \label{fact-curve}
		The curve $\xi(A)$ has the following properties. \\
		\textnormal{\bf (1)} If $\sigma$ and $\sigma'$ are two pieces of $\xi(A)$ contributed by $a$ and $a'$ respectively, then $\sigma$ is to the left of $\sigma'$ on $\xi(A)$ iff $a$ is to the left of $a'$. \\
		\textnormal{\bf (2)} For each $a \in A$, there is at most one piece of $\xi(A)$ contributed by $a$.
		Therefore, the complexity of $\xi(A)$, i.e., the number of the pieces, is $O(|A|)$.
		%\textnormal{\bf (3)} $\xi(W)$ can be computed in $O(|W| \cdot \log |W|)$ time, provided the points in $W$.
	\end{fact}
	\textit{Proof.}
	We first notice that the property \textbf{(2)} follows directly from the property \textbf{(1)}.
	Indeed, if there are two pieces $\sigma_1$ and $\sigma_2$ both contributed by $a \in A$ where $\sigma_1$ is to the left of $\sigma_2$ on $\xi(A)$, they must be non-adjacent (otherwise they become one piece).
	Let $\sigma'$ be a piece of $\xi(A)$ in between $\sigma_1$ and $\sigma_2$ and assume $\sigma'$ is contributed to $a' \in A$.
	Since $\sigma'$ is to the right of $\sigma_1$, $a'$ is to the right of $a$ by the property \textbf{(1)}.
	On the other hand, since $\sigma'$ is to the left of $\sigma_2$, $a'$ is to the left of $a$ by the property \textbf{(1)}, which is a contradiction.
	Now it suffices to show the property \textbf{(1)}.
	Let $\sigma$ and $\sigma'$ be two pieces of $\xi(A)$ contributed by $a$ and $a'$ respectively.
	Assume that $a$ is to the left of $a'$.
	If $A = \{a,a'\}$, then the fact that $\sigma$ is to the left of $\sigma'$ can be easily verified by elementary geometry.
	Otherwise, let $A_0 = \{a,a'\}$ and $\sigma_0$ (resp, $\sigma_0'$) be the piece of $\xi(A_0)$ contributed by $a$ (resp., $a'$).
	We know that $\sigma_0$ is to the left of $\sigma_0'$.
	%Assume $\sigma$ is to the left of $\sigma'$ on $\xi(A)$ and $a$ is to the right of $a'$.
	Since $A_0 \subseteq A$, we have $\sigma \subseteq \sigma_0$ and $\sigma' \subseteq \sigma_0'$.
	As such, $\sigma$ is to the left of $\sigma'$.
	\hfill $\Box$
	\smallskip
	
\begin{figure}[t]
		\begin{minipage}[t]{\linewidth}
			\begin{center}
				\includegraphics[totalheight=0.9in]{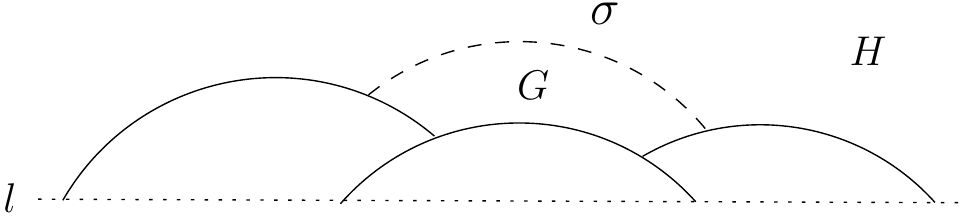}
				\caption{\footnotesize
					Illustrating the new piece $\sigma$ (the dashed curve) defined by $u_i$. The upper envelope of the solid curves is $\xi_{i-1}$. The upper envelope of all the curves (solid and dashed) is $\xi_i$.}
				\label{fig:face}
			\end{center}
		\end{minipage}
	\end{figure}

	\noindent
	The above observation in fact implies the linear complexity of $\varPhi$.
	Define $U_i = \{u_1,\dots,u_i\}$ and $\xi_i = \xi(U_i)$ for $i \in \{1,\dots,|U|\}$.
	Note that $\xi_i$ is the boundary of (the closure of) $F_i$.
	\begin{corollary} \label{cor-linear}
		We have $|\varPhi| = O(|U|)$.
		Furthermore, the subdivision $\varPhi$ has at most one face labeled as $i$ for all $i \in \{1,\dots,|U|\}$.
		%For all $i \in \{1,\dots,|U|\}$, $\varPhi_i$ has at most two more vertices than $\varPhi_{i-1}$.
	\end{corollary}
	\textit{Proof.}
	Let $\mathcal{G}_i$ be the set of the inner faces of $\varPhi_i$ (i.e., the faces other than the outer face $F_i$).
	Then $\mathcal{G}_0 \subseteq \cdots \subseteq \mathcal{G}_{|U|}$.
	Note that the faces of $\varPhi$ labeled as $i$ are exactly those in $\mathcal{G}_i \backslash \mathcal{G}_{i-1}$.
	Fix $i \in \{1,\dots,|U|\}$.
	We shall show that \textbf{(1)} $|\mathcal{G}_i \backslash \mathcal{G}_{i-1}| \leq 1$ and \textbf{(2)} $\varPhi_i$ has at most two more vertices than $\varPhi_{i-1}$.
	Note that \textbf{(2)} implies $|\varPhi| = O(|U|)$.
	If $F_{i-1} \cap \odot_{u_i} = \emptyset$, then $\varPhi_i = \varPhi_{i-1}$.
	In this case, $\mathcal{G}_i = \mathcal{G}_{i-1}$ and $\varPhi_i$ has the same number of vertices as $\varPhi_{i-1}$.
	So assume $F_{i-1} \cap \odot_{u_i} \neq \emptyset$.
	In this case, at least one piece of $\xi_i$ is contributed by $u_i$.
	Further applying the property \textbf{(2)} of Fact~\ref{fact-curve}, we know that there is exactly one piece $\sigma$ of $\xi_i$ contributed by $u_i$ (e.g., see Fig.~\ref{fig:face}).
	
	The portion of $\xi_i$ to the left (resp., right) of $\sigma$ agrees with $\xi_{i-1}$.
	Now the area above $\xi_i$ (resp., $\xi_{i-1}$) is $F_i$ (resp., $F_{i-1}$), and the area in between $\xi_{i-1}$ and $\xi_i$ consists of the new inner faces in $\varPhi_i$ (i.e., those in $\mathcal{G}_i \backslash \mathcal{G}_{i-1}$).
	Since $\sigma$ is the only piece of $\xi_i$ contributed by $u_i$, the area in between $\xi_{i-1}$ and $\xi_i$ is in fact a connected region $G$ whose upper boundary is $\sigma$ and lower boundary is a portion of $\xi_{i-1}$ sharing the same left/right endpoints with $\sigma$ (e.g., see Fig.~\ref{fig:face}).
	%By construction, $\varPhi_i$ is obtained from $\varPhi_{i-1}$ by decomposing $F_{i-1}$ (the area above $\xi_{i-1}$) into $F_i$ (the area above $\xi_i$) and $G$.
	So $\mathcal{G}_i \backslash \mathcal{G}_{i-1} = \{G\}$.
	%Hence, $\varPhi_i$ has one more face than $\varPhi_{i-1}$.
	Also, $\varPhi_i$ has two more vertices than $\varPhi_{i-1}$, which are the left and right endpoints of $\sigma$.
	Therefore, $|\varPhi_i| - |\varPhi_{i-1}| = O(1)$ and $|\varPhi| = O(|U|)$.
	%We notice that $\partial \odot_{u_i} \cap F_{i-1}$ only has one connected component, where $\partial \odot_{u_i}$ is the boundary of $\odot_{u_i}$.
	%Indeed, each connected component of $\partial \odot_{u_i} \cap F_{i-1}$ corresponds to a piece of $\xi_i$ contributed by $u_i$.
	%By the property \textbf{(2)} of Fact~\ref{fact-curve}, there is at most one piece of $\xi_i$ contributed by $u_i$. This implies that $\partial \odot_{u_i} \cap F_{i-1}$ only has one connected component.
	%Therefore, $\partial \odot_{u_i} \cap F_{i-1}$ is a curve which decomposes $F_{i-1}$ into two connected regions of which one is $F_{i-1} \backslash \odot_{u_i}$ and the other is $F_{i-1} \cap \odot_{u_i}$.
	%By construction, $F_{i-1} \backslash \odot_{u_i}$ becomes the outer face of $\varPhi_i$ and $F_{i-1} \cap \odot_{u_i}$ becomes a new face of $\varPhi_i$ labeled as $i$.
	\hfill $\Box$
	\smallskip
	
	Next, we show how to construct the subdivision $\varPhi$ efficiently.
	Our algorithm for constructing $\varPhi$ has $|U|$ iterations.
	In the $i$-th iteration, we shall compute the face of $\varPhi$ labeled as $i$ (if it exists).
	To this end, we need to maintain the curve $\xi_i$.
	Naturally, such an $x$-monotone curve can be stored in a binary search tree in which the nodes are one-to-one corresponding to the pieces in left-right order.
	So we use a (balanced) BST $\mathcal{T}$ to maintain $\xi_i$, that is, we guarantee that the curve stored in $\mathcal{T}$ is $\xi_i$ when the $i$-th iteration is done.
	Note that the number of the nodes in $\mathcal{T}$ is always $O(|U|)$ by the property \textbf{(2)} of Fact~\ref{fact-curve}.
	Suppose we are now at the beginning of the $i$-th iteration and the curve stored in $\mathcal{T}$ is $\xi_{i-1}$.
	We need to update the curve in $\mathcal{T}$ to $\xi_i$ and at the same time compute the face labeled as $i$ in the $i$-th iteration.
	To this end, a critical step is to find the piece $\sigma$ of $\xi_i$ contributed by $u_i$ or decide that $\sigma$ does not exist.
	It suffices to find the left and right endpoints of $\sigma$, which are the two intersection points of $\xi_{i-1}$ and the boundary of $\odot_{u_i}$.
	We can find these endpoints by binary search on $\xi_{i-1}$ as follows.
	
	Suppose we want to find the left endpoint of $\sigma$, say $p$.
	Let $\sigma'$ be a piece of $\xi_{i-1}$.
	Also, let $p'$ and $q'$ be the left and right endpoints of $\sigma'$ respectively.
	%If $\sigma'$ is entirely contained in $\odot_{u_i}$, then $p$ must be to the left of $\sigma'$.
	If $p' \in \odot_{u_i}$, then $p$ must be to the left of $p'$ and hence $p$ lies on some piece of $\xi_{i-1}$ to the left of $\sigma'$.
	If $p' \notin \odot_{u_i}$ and $q' \in \odot_{u_i}$, then $p$ is the intersection point of $\sigma'$ and the boundary of $\odot_{u_i}$.
	The remaining case is that $p',q' \notin \odot_{u_i}$.
	In this case, assume $\sigma'$ is contributed by $u_j$ for some $j<i$.
	Since $\sigma'$ is a portion of the unit disk $\odot_{u_j}$ and $u_j,u_i \in \Box$, the fact $p',q' \notin \odot_{u_i}$ implies that $\sigma' \cap \odot_{u_i} = \emptyset$.
	Thus, $\sigma'$ is also a piece of $\xi_i$.
	If $u_i$ is to the left (resp., right) of $u_j$, then $\sigma$ is to the left (resp., right) of $\sigma'$ by the property \textbf{(1)} of Fact~\ref{fact-curve}, so is $p$.
	In sum, given a piece $\sigma'$ of $\xi_{i-1}$, we can know in constant time whether $p$ is on $\sigma'$ or to the left/right of $\sigma'$; furthermore, if $p$ is on $\sigma'$, it can be directly computed.
	With this observation in hand, $p$ can be computed (if it exists) in $O(\log |U|)$ time by searching in $\mathcal{T}$.
	Using the same method, we can also find the right endpoint $q$ of $\sigma$.
	
	We then use $\mathcal{T}$ to report all the pieces of $\xi_{i-1}$ in between $p$ and $q$ in left-right order, which we denote by $\sigma_1,\dots,\sigma_d$.
	This takes $O(\log |U| + d)$ time.
	Let $\sigma_0$ and $\sigma_{d+1}$ be the piece containing $p$ and $q$, respectively.
	Since $\xi_i$ is obtained from $\xi_{i-1}$ by replacing the portion in between $p$ and $q$ with $\sigma$, we can update $\mathcal{T}$ by deleting $\sigma_1,\dots,\sigma_d$, adding $\sigma$ to $\mathcal{T}$, and modifying $\sigma_0$ and $\sigma_{d+1}$.
	After this, the curve stored in $\mathcal{T}$ is updated to $\xi_i$.
	With $\sigma_0,\dots,\sigma_{d+1}$ and $\sigma$ in hand, to compute the face of $\varPhi$ labeled as $i$ is also easy.
	As we see in the proof of Corollary~\ref{cor-linear}, the face labeled as $i$ is just the region whose upper boundary is $\sigma$ and lower boundary is the portion of $\xi_{i-1}$ in between $p$ and $q$.
	Thus, using the pieces $\sigma_0,\dots,\sigma_{d+1}$ and $\sigma$, the face labeled as $i$ can be computed in $O(d+1)$ time.
	The time cost for the $i$-th iteration is $O(\log |U| + d)$.
	Note that $d$ is at most the number of the edges of the face labeled as $i$.
	The overall time for all the $|U|$ iterations is $O(|U| \cdot \log |U| + \sum_{i=1}^{|U|} d_i)$, where $d_i$ is the number of the edges of the face labeled as $i$.
	Since $|\varPhi| = O(|U|)$, we know that $\sum_{i=1}^{|U|} d_i = O(|U|)$ and hence the $|U|$ iterations take $O(|U| \cdot \log |U|)$ time in total.
	At the end of the last iteration, all the faces of $\varPhi$ labeled as $1,\dots,|U|$ are computed and the curve stored in $\mathcal{T}$ is $\xi_{|U|}$.
	Finally, we compute the outer face of $\varPhi$ (i.e., $F_{|U|}$) by recovering the curve $\xi_{|U|}$ via an in-order traversal in $\mathcal{T}$ (note that $\xi_{|U|}$ is the boundary of $F_{|U|}$).
	In this way, the subdivision $\varPhi$ is constructed in $O(|U| \cdot \log |U|)$.
	
	Once we obtain $\varPhi$, we can build an optimal point-location data structure on $\varPhi$ in $O(|U|)$ time~\cite{EdelsbrunnerOp86}, since $|\varPhi| = O(|U|)$.
	As argued before, we can use this data structure to compute $i(v)$ for all $v \in V$ above the top bounding line $l$ of $\Box$.
	By building similar data structures for the bottom/left/right bounding lines of $\Box$, we can compute $i(v)$ for all $v \in V$.
	The overall running time, including the time for constructing the data structures and answering point-location queries, is $O(k \log k)$ where $k = |U|+|V|$.
	
	Now we consider the case where the points in $U$ do not necessarily lie in a grid cell.
	Let $\mathcal{U}$ be the collection of the cells in $\varGamma$ containing at least one point of $U$.
	For each cell $\Box \in \mathcal{U}$, we build in $O(|U_\Box| \cdot \log |U_\Box|)$ time a data structure described above, denoted by $\mathcal{D}(\Box)$, which can report $\min\{i: u_i \in \Box \text{ and } v \in \odot_{u_i}\}$ for any point $v \in V$ in $O(\log |U_\Box|)$ time.
	The total time for building these data structures is then $O(|U| \cdot \log |U|)$.
	Using these data structures, the simplest way to compute $i(v)$ for a point $v \in V$ is to query $\mathcal{D}(\Box)$ for all $\Box \in \mathcal{U}$ and take the minimum of the reported values.
	However, we need to query $|\mathcal{U}|$ data structures for each $v \in V$, which takes too much time as $|\mathcal{U}| = |U|$ in the worst case.
	To resolve this issue, we notice that in order to compute $i(v)$ for a point $v \in V$, we only need to query $\mathcal{D}(\Box)$ for the cells $\Box \in \mathcal{U}$ that are contained in the patch $\boxplus_v$ (the number of which is at most 25), because $v \notin \odot_u$ for any $u \notin \boxplus_v$.
	Therefore, each point in $V$ requires at most 25 queries; the total time for considering all points in $V$ is $O(|V| \cdot \log |U|)$.
	It follows that the first loop of Algorithm~\ref{alg-refine} takes $O(k \log k)$ time where $k = |U|+|V|$.
	%Therefore, \textsc{Update}$(U,V)$ can be done in $O(f(k)+k \log k)$ time.
	
	\subsection{Putting Everything Together}
	As argued before, except the two \textsc{Update} sub-routines, Algorithm~\ref{alg-SSSP} runs in $O(n \log n)$ time.
	Section~\ref{sec-correction} shows that the first update can be done in $O(|S_{\boxplus_c}| \cdot \log n)$ time.
	Section~\ref{sec-refine} demonstrates that the second update of each iteration can be done in $O(f(k) + k \log k)$ time where $k = |A_{\Box_c}| + |A_{\boxplus_c}| = O(|S_{\boxplus_c}|)$ and $f(k)$ is the time for solving the OIWNN problem with $k$ operations.
	%Assume the function $f$ satisfies $f(a+b) \geq f(a)+f(b)$ and for any $c > 0$ there exists $c' > 0$ such that $f(cn) \leq c'f(n)$ for all $n>0$.
	Noting the fact $\sum_{i=1}^m |S_{\boxplus_{c_i}}| \leq 25 n$, we can conclude the following.
	\begin{theorem}
		Suppose the OIWNN problem with $k$ operations can be solved in $f(k)$ time, where $f(k)/k$ is a non-decreasing function.
		Then there exists an SSSP algorithm in weighted unit-disk graphs with $O(n \log n + f(n))$ running time, where $n$ is the number of the vertices.
	\end{theorem}
	\textit{Proof.}
	According to our analysis and the fact $\sum_{i=1}^m |S_{\boxplus_{c_i}}| \leq 25 n$, the overall time of Algorithm~\ref{alg-SSSP} is $O(n \log n + \sum_{i=1}^m f(|S_{\boxplus_{c_i}}|))$.
	Since $f(k)/k$ is non-decreasing, we have
	\begin{equation*}
	\sum_{i=1}^m f(|S_{\boxplus_{c_i}}|) = \sum_{i=1}^m |S_{\boxplus_{c_i}}| \cdot \frac{f(|S_{\boxplus_{c_i}}|)}{|S_{\boxplus_{c_i}}|} \leq \sum_{i=1}^m |S_{\boxplus_{c_i}}| \cdot \frac{f(n)}{n} \leq 25 f(n).
	\end{equation*}
	Therefore, Algorithm~\ref{alg-SSSP} runs in $O(n \log n + f(n))$ time.
	\hfill $\Box$
	\medskip
	
	\noindent
	%Using the standard logarithmic method~\cite{ref:BentleyDe79} (even for online operations), we can immediately obtain $f(k)=O(k\log^2 k)$ (using $O(k)$ space).
	Using the standard logarithmic method \cite{ref:BentleyDe79} (see also \cite{deBerg2016fine} with an additional ``bulk update'' operation), we can solve the OIWNN problem (even the online version) with $k$ operations in $O(k \log^2 k)$ time using linear space, implying $f(k) = O(k \log^2 k)$.
	To explore the offline nature of our OIWNN problem, we give in Appendix~\ref{appx-WNN} an easier solution with the same performance. 
	%linear-space algorithm that solves in $O(k \log k)$ time the OIWNN problem with $k$ operations, which implies $f(k) = O(k \log^2 k)$.
	By plugging in this algorithm, 
	%we can implement Algorithm~\ref{alg-SSSP} in $O(n \log^2 n)$ time and linear space, 
	we obtain the following corollary.
	\begin{corollary} \label{cor-exact}
		There exists an SSSP algorithm in weighted unit-disk graphs with $O(n \log^2 n)$ time and $O(n)$ space, where $n$ is the number of the vertices.
	\end{corollary}
	
	\section{The Approximation Algorithm} \label{sec-approx}
	We now modify our algorithm framework in the last section (Algorithm~\ref{alg-SSSP}) to obtain a $(1+\varepsilon)$-approximate 
	algorithm for any $\varepsilon>0$.
	Again, let $(S,s)$ be the input of the problem where $|S| = n$ and $G$ be the weighted unit-disk graph induced by $S$.
	%Let $(S,s)$ be the input where $S$ is a set of $n$ points in the plane and $s \in S$ is the source point.
	Formally, a $(1+\varepsilon)$-approximate algorithm computes
	%, for an input $(S,s)$ where $S$ is a set of $n$ points in the plane and $s \in S$ is the source point, 
	two tables $\text{dist}[\cdot]$ and $\text{pred}[\cdot]$ indexed by the points in $S$ such that $\text{dist}[a] \leq (1+\varepsilon) \cdot d_G(s,a)$ and $\text{dist}[a] = \text{dist}[\text{pred}[a]]+\lVert \text{pred}[a]-a \rVert$ for all $a \in S$.
	Note that the two tables $\text{dist}[\cdot]$ and $\text{pred}[\cdot]$ enclose, for each point $a \in S$, a path from $s$ to $a$ in $G$ that is a $(1+\varepsilon)$-approximation of the shortest path from $s$ to $a$.
	
	Our algorithm is shown in Algorithm~\ref{alg-approxSSSP}, which differs from our exact algorithm (Algorithm~\ref{alg-SSSP}) as follows.
	First, in the initialization, we directly compute the dist-values and pred-values of all the neighbors of $s$ in $G$ (line~4-6); note that if $a$ is a neighbor of $s$ then the shortest path from $s$ to $a$ is $\langle s,a \rangle$, because $G$ is a weighted unit-disk graph.
	Second, the first update in Algorithm~\ref{alg-SSSP} is replaced with two update procedures (line~10-11).
	Finally, the second update in Algorithm~\ref{alg-SSSP} is replaced with an approximate update (line~12) in Algorithm~\ref{alg-approxSSSP}, which involves a new sub-routine \textsc{ApproxUpdate} defined as follows.
	If $U$ and $V$ are two \textit{disjoint} subsets of $S$,  \textsc{ApproxUpdate}$(U,V)$ conceptually does the following.
	\begin{enumerate}
		%\item $\text{dist}'[u] \leftarrow \text{dist}[u]$ for all $u \in U$.
		\item For each $v \in V$, pick a point $p_v \in U \cap \odot_v$ such that $\text{dist}[p_v] + \lVert p_v-v \rVert \leq \text{dist}[u] + \lVert u-v \rVert + \varepsilon/2$ for all $u \in U \cap \odot_v$.
		\item For all $v \in V$, if $\text{dist}[v] < \text{dist}[p_v] + \lVert p_v-v \rVert$, then
		$\text{dist}[v] \leftarrow \text{dist}[p_v] + \lVert p_v-v \rVert$ and $\text{pred}[v] \leftarrow p_v$.
	\end{enumerate}
	Unlike the \textsc{Update} sub-routine, \textsc{ApproxUpdate} cannot result in data inconsistency because we require $U$ and $V$ to be disjoint.
	
	\begin{algorithm}[tbph]
		\caption{\textsc{ApproxSSSP}$(S,s)$}
		\begin{algorithmic}[1]
			\State $\text{dist}[a] \leftarrow \infty$ for all $a \in S$
			\State $\text{pred}[a] \leftarrow \text{NIL}$ for all $a \in S$
			\State $\text{dist}[s] \leftarrow 0$
			\For{$a \in (S \backslash \{s\}) \cap \odot_s$}
			\State $\text{dist}[a] \leftarrow \lVert s-a \rVert$
			\State $\text{pred}[a] \leftarrow s$        
			\EndFor
			\State $A \leftarrow S$
			%\State $I \leftarrow S \backslash \{s\}$
			\While{$A \neq \emptyset$} \Comment{Main loop}
			\State $c \leftarrow \arg \min_{a \in A} \{\text{dist}[a]\}$
			\State \textsc{Update}$(A_{\boxplus_c} \backslash A_{\Box_c},A_{\Box_c})$
			\State \textsc{Update}$(A_{\Box_c},A_{\Box_c})$
			\State \textsc{ApproxUpdate}$(A_{\Box_c},A_{\boxplus_c} \backslash A_{\Box_c})$ \Comment{Approximate update}
			\State $A \leftarrow A \backslash A_{\Box_c}$
			\EndWhile
			\State \textbf{return} $\text{dist}[\cdot]$ and $\text{pred}[\cdot]$
			%\nolinenumbers
		\end{algorithmic}
		\label{alg-approxSSSP}
	\end{algorithm}
	
	The basic idea of Algorithm~\ref{alg-approxSSSP} is similar to that of our exact  algorithm.
	To verify the correctness of the algorithm, we need to introduce some notations.
	For $a \in S$, let $l_a$ be the number of the edges on the path $\pi_G(s,a)$ and define $\tau_a = d_G(s,a) + (l_a-1) \cdot (\varepsilon/2)$.
	Also, as in Section~\ref{sec-exact}, we use $\lambda_a$ to denote the $s$-predecessor of $a$.
	We first notice the following fact.
	\begin{fact} \label{fact-approx}
		For all $a \in S$, $\tau_a \leq (1+\varepsilon) \cdot d_G(s,a)$.
	\end{fact}
	\textit{Proof.}
	Suppose $\pi_G(s,a) = \langle z_0,z_1,\dots,z_{l_a} \rangle$ where $z_0 = s$ and $z_{l_a} = a$.
	Note that $\lVert z_i - z_{i+2} \rVert > 1$ for all $i \in \{0,\dots,l_a-2\}$, for otherwise $\langle z_0,z_1,\dots,\hat{z}_{i+1},\dots,z_{l_a} \rangle$ would be a shorter path from $s$ to $a$ than $\pi_G(s,a)$ (here $\hat{z}_{i+1}$ means $z_{i+1}$ is absent in the sequence).
	Therefore, $d_G(s,a) \geq (l_a-1)/2$, and $\tau_a = d_G(s,a) + (l_a-1) \cdot (\varepsilon/2) \leq (1+\varepsilon) \cdot d_G(s,a)$.
	\hfill $\Box$
	\medskip
	
	\noindent
	Let $m$ be the number of iterations of the main loop and $c_i$ be the point $c$ picked in the $i$-th iteration.
	Note that Fact~\ref{fact-disjoint} also holds for Algorithm~\ref{alg-approxSSSP}.
	Further, we have the following observation, which is similar to Lemma~\ref{lem-correct} in Section~\ref{sec-exact}.
	\begin{lemma} \label{lem-correct'}
		Algorithm~\ref{alg-approxSSSP} has the following properties. \\
		\textnormal{\bf (1)} When the $i$-th iteration begins, $\textnormal{dist}[a] \leq \tau_a$ for all $a \in S$ with $\tau_a \leq \textnormal{dist}[c_i]$. \\
		\textnormal{\bf (2)} After line~11 of the $i$-th iteration, $\textnormal{dist}[a] \leq \tau_a$ for all $a \in S_{\Box_{c_i}}$. \\
		\textnormal{\bf (3)} When the $i$-th iteration ends, $\textnormal{dist}[a] \leq \tau_a$ for all $a \in S$ with $\lambda_a \in S_{\Box_{c_i}}$.
	\end{lemma}
	\textit{Proof.}
	We notice that the property \textbf{(3)} follows from the property \textbf{(2)}.
	To see this, let $a \in S$ be a point such that $\lambda_a = r \in S_{\Box_{c_i}}$.
	(Recall that $\lambda_a$ is the $s$-predecessor of $a$.)
	Then $a \in S_{\boxplus_{c_i}}$.
	The property \textbf{(2)} implies $\text{dist}[r] \leq \tau_r$ after line~11 of the $i$-th iteration.
	If $a \in A_{\Box_{c_i}}$, then the fact that $\text{dist}[a] \leq \tau_a$ at the end of the $i$-th iteration directly follows from the property \textbf{(2)}.
	If $a \in A_{\boxplus_{c_i}} \backslash A_{\Box_{c_i}}$, then the approximate update makes $\text{dist}[a] \leq \text{dist}[r] + \lVert r-a \rVert + \varepsilon/2$.
	By definition, $l_a = l_r+1$ and
	\begin{equation*}
	\tau_a = d_G(s,a) + (l_a-1) \cdot (\varepsilon/2) = d_G(s,r) + \lVert r-a \rVert + l_r \cdot (\varepsilon/2) = \tau_r + \lVert r-a \rVert + \varepsilon/2.
	\end{equation*}
	Because $\text{dist}[r] \leq \tau_r$, we obtain $\text{dist}[a] \leq \tau_a$ at the end of the $i$-th iteration.
	If $a \in S_{\boxplus_{c_i}} \backslash A_{\boxplus_{c_i}}$, then $a \in A_{\Box_{c_j}}$ for some $j<i$ (since $a$ got removed from $A$ in some previous iteration) and the property \textbf{(2)} guarantees that $\text{dist}[a] \leq \tau_a$ after line~11 of the $j$-th iteration.
	
	As such, we only need to verify the first two properties.
	We achieve this using induction on $i$.
	The base case is $i=1$.
	Note that $c_1 = s$ and $\text{dist}[s] = 0$ at the beginning of the first iteration.
	Thus, to see the property \textbf{(1)}, we only need to guarantee that $\text{dist}[s] \leq \tau_s = 0$ when the first iteration begins, which is clearly true.
	Also, the property \textbf{(2)} clearly holds.
	Indeed, the initialization already set $\text{dist}[a]$ to $d_G(s,a)$ for all $a \in A_{\Box_s}$, since $A_{\Box_s} \subseteq \odot_s$.
	%After line~11 of the first iteration, we have $\text{dist}[a] = \lVert s-a \rVert = d_G(s,a)$ for all $a \in S_{\Box_{c_1}}$, hence the property \textbf{(2)} is satisfied.
	Assume the lemma holds for all $i < k$, and we show it also holds in the $k$-th iteration.
	
	To see the property \textbf{(1)}, consider the moment when the $k$-th iteration begins.
	Let $a \in S$ be a point such that $\tau_a \leq \text{dist}[c_i]$ at that time.
	Assume for a contradiction that $\text{dist}[a] > \tau_a$.
	Suppose $\pi_G(s,a) = \langle z_0,z_1,\dots,z_t \rangle$ where $z_0 = s$ and $z_t = a$.
	Define $j$ as the largest index such that $\text{dist}[z_j] \leq \tau_{z_j}$.
	Note that $j \in \{0,\dots,t-1\}$ because $\text{dist}[s] \leq \tau_s = 0$ and $\text{dist}[a] > \tau_a$.
	We have $\tau_{z_j} < \tau_a$ since $d_G(s,z_j)<d_G(s,a)$ and $l_{z_j} < l_a$.
	Therefore, $\text{dist}[z_j] \leq \tau_{z_j} < \tau_a \leq \text{dist}[c_k]$.
	This implies $z_j \notin A$ (otherwise it contradicts the fact that $c_k$ is the point in $A$ with the smallest dist-value).
	%Also, $z_j \notin I$ as $\text{dist}[z_j] < \infty$.
	It follows $z_j \in S_{\Box_{c_i}}$ for some $i<k$, as it got removed from $A$ in some previous iteration.
	Then by our induction hypothesis and the property \textbf{(3)}, we have $\text{dist}[z_{j+1}] \leq \tau_{z_{j+1}}$ at the end of the $i$-th iteration (and thus at the beginning of the $k$-th iteration), because $\lambda_{z_{j+1}} = z_j$.
	However, this contradicts the fact that $\text{dist}[z_{j+1}] > \tau_{z_{j+1}}$.
	As such, $\text{dist}[a] \leq \tau_a$ holds when the $k$-th iteration begins.
	
	Next, we prove the property \textbf{(2)}.
	For convenience, in what follows, we use $A$ to denote the set $A$ during the $k$-th iteration (before line~13).
	We have $S_{\Box_{c_k}} = A_{\Box_{c_k}}$, since $A = S \backslash (\bigcup_{i=1}^{k-1} S_{\Box_{c_i}})$ and $c_k \notin \Box_{c_i}$ for all $i<k$ by Fact~\ref{fact-disjoint}.
	Let $a \in A_{\Box_{c_k}}$ be a point and $r = \lambda_a$.
	We want to show that $\text{dist}[a] \leq \tau_a$ after line~11 of the $k$-th iteration.
	If $r \notin A$, then $r$ got removed from $A$ in the $i$-th iteration for some $i<k$, i.e., $r \in S_{\Box_{c_i}}$.
	By our induction hypothesis and the property \textbf{(3)}, $\text{dist}[a] \leq \tau_a$ at the end of the $i$-th iteration and hence in all the next iterations.
	So assume $r \in A$.
	%If $r = s$, then by our induction hypothesis and the property \textbf{(3)}, $\text{dist}[a] \leq \tau_a$ at the end of the first iteration.
	Let $r' = \lambda_r$ (e.g., see Fig.~\ref{fig:predecessor}).
	Note that $\lVert r' - a \rVert > 1$, otherwise the path $\pi_G(s,r') \circ \langle r',a \rangle$ is shorter than $\pi_G(s,r') \circ \langle r',r,a \rangle = \pi_G(s,a)$, contradicting the fact that $\pi_G(s,a)$ is the shortest path from $s$ to $a$.
	It follows that
	\begin{equation*}
	d_G(s,a) = d_G(s,r') + d_G(r',a) \geq d_G(s,r') + \lVert r'-a \rVert > d_G(s,r') + 1,
	\end{equation*}
	and hence $\tau_a > \tau_{r'} + 1$ (for $l_a > l_{r'}$).
	With this observation in hand, we consider two cases separately: $\tau_{r'} \geq \text{dist}[c_k]$ and $\tau_{r'} \leq \text{dist}[c_k]$ at the beginning of the $k$-th iteration.
	
	\begin{itemize}
		\item 
		Assume $\tau_{r'} \geq \text{dist}[c_k]$ at the beginning of the $k$-th iteration.
		Since $c_k$ is the point in $A$ with the smallest dist-value, line~10 and 11 do not change $\text{dist}[c_k]$.
		As such, after line~11 of the $k$-th iteration, we have
		\begin{equation*}
		\text{dist}[a] \leq \text{dist}[c_k] + \lVert c_k-a \rVert \leq \tau_{r'} + 1 < \tau_a,
		\end{equation*}
		because $a$ and $c_k$ are both in ${\Box_{c_k}}$.
		
		\item
		Assume $\tau_{r'} < \text{dist}[c_k]$ at the beginning of the $k$-th iteration.
		Then by the property \textbf{(1)}, we have $\text{dist}[r'] \leq \tau_{r'} < \text{dist}[c_k]$ when the $k$-th iteration begins.
		This implies $r' \notin A$, because $c_k$ is the point in $A$ with the smallest dist-value.
		Therefore, $r'$ got removed from $A$ in the $i$-th iteration for some $i<k$, i.e., $r' \in S_{\Box_{c_i}}$.
		By our induction hypothesis and the property \textbf{(3)}, $\text{dist}[r] \leq \tau_r$ at the end of the $i$-th iteration (and hence in all the next iterations).
		Note that $r \in \boxplus_{c_k}$, because $r \in \odot_a$.
		We further have $r \in A_{\boxplus_{c_k}}$, as we assumed $r \in A$.
		Define $\gamma$ as the dist-value of $r$ just before line~10 of the $k$-th iteration (we have $\gamma \leq \tau_r$ because $\text{dist}[r] \leq \tau_r$ at the end of the $i$-th iteration).
		Then after line~10 and 11 of the $k$-th iteration, we have
		\begin{equation*}
		\text{dist}[a] \leq \gamma + \lVert r-a \rVert \leq \tau_r + \lVert r-a \rVert \leq \tau_a,
		\end{equation*}
		where the last inequality is due to that $d_G(s,r)=d_G(s,a) + \lVert r-a \rVert$ and $l_r < l_a$.
	\end{itemize}
	This completes the proof of the property \textbf{(2)} and also the entire lemma.
	\hfill $\Box$
	\medskip
	
	\noindent
	By the above lemma, we see that $\text{dist}[a] \leq \tau_a$ for all $a \in S$ at the end of Algorithm~\ref{alg-approxSSSP}.
	Indeed, any point $a \in S$ belongs to $S_{\Box_{c_i}}$ for some $i \in \{1,\dots,m\}$, thus the property \textbf{(2)} of Lemma~\ref{lem-correct'} guarantees that $\text{dist}[a] \leq \tau_a$.
	Using Fact~\ref{fact-approx}, we further conclude that $\text{dist}[a] \leq (1+\varepsilon) \cdot d_G(s,a)$ for all $a \in S$ at the end of Algorithm~\ref{alg-approxSSSP}.
	Next, we need to check the correctness of the $\text{pred}[\cdot]$ table.
	We want $\text{dist}[a] = \text{dist}[\text{pred}[a]] + \lVert \text{pred}[a]-a \rVert$ for all $a \in S$.
	As mentioned in Section~\ref{sec-exact}, the procedure \textsc{Update}$(U,V)$ may result in data inconsistency, namely $\text{dist}[v] > \text{dist}[\text{pred}[v]] + \lVert \text{pred}[v]-v \rVert$ for some $v \in V$, when $U$ and $V$ are not disjoint.
	In Algorithm~\ref{alg-approxSSSP}, the only place where this can happen is line~11 (note that the \textsc{Update} sub-routine in line~10 acts on two disjoint sets).
	However, the following lemma shows that even line~11 cannot result in data inconsistency.
	\begin{lemma} \label{lemma-consistency}
		After line~11 of each iteration in Algorithm~\ref{alg-approxSSSP}, we have $\textnormal{dist}[a] \leq \textnormal{dist}[b] + \lVert b-a \rVert$ for all $a,b \in A_{\Box_c}$.
		In particular, at any moment of Algorithm~\ref{alg-approxSSSP}, we always have $\textnormal{dist}[a] = \textnormal{dist}[\textnormal{pred}[a]] + \lVert \textnormal{pred}[a]-a \rVert$ for all $a \in S$.
	\end{lemma}
	\textit{Proof.}
	For a point $a \in A_{\Box_c}$, let $\text{dist}'[a]$ be the dist-value of $a$ just before line~11.
	Consider two points $a,b \in A_{\Box_c}$.
	Since $b \in \odot_a$, we have $\text{dist}[a] \leq \text{dist}'[b] + \lVert b-a \rVert$ after line~11 by the definition of the \textsc{Update} sub-routine.
	If $\text{dist}[b] = \text{dist}'[b]$ after line~11, we are done.
	Assume $\text{dist}[b] < \text{dist}'[b]$ after line~11.
	Then $\text{pred}[b] \in A_{\Box_c}$ and $\text{dist}[b] = \text{dist}'[\text{pred}[b]] + \lVert \text{pred}[b]-b \rVert$ after line~11, because $\text{dist}[b]$ and $\text{pred}[b]$ are changed during the procedure \textsc{Update}$(A_{\Box_c},A_{\Box_c})$.
	We write $p = \text{pred}[b]$.
	Since $p \in A_{\Box_c}$, after line~11 we have
	\begin{equation*}
	\text{dist}[a] \leq \text{dist}'[p] + \lVert p-a \rVert \leq \text{dist}'[p] + \lVert p-b \rVert + \lVert b-a \rVert = \text{dist}[b] + \lVert b-a \rVert,
	\end{equation*}
	which proves the first statement of the lemma.
	
	Note that the first statement implies that no data inconsistency can happen in line~11.
	To see this, consider a point $a \in A_{\Box_c}$.
	Assume $\text{dist}[a] = \text{dist}[\text{pred}[a]] + \lVert \text{pred}[a]-a \rVert$ before line~11.
	We claim that this equation still holds after line~11.
	If $\text{pred}[a] \notin A_{\Box_c}$ after line~11, then $\text{dist}[a]$ and $\text{pred}[a]$ do not change during line~11, and hence the equation holds.
	If $\text{pred}[a] \in A_{\Box_c}$ after line~11, then the first statement of the lemma guarantees that the equation holds after line~11 (because at any moment of the algorithm, $\text{dist}[a] \geq \text{dist}[\text{pred}[a]] + \lVert \text{pred}[a]-a \rVert$ always holds).
	Therefore, line~11 cannot result in data inconsistency.
	It follows that $\text{dist}[a] = \text{dist}[\text{pred}[a]] + \lVert \text{pred}[a]-a \rVert$ for all $a \in S$ at any moment of Algorithm~\ref{alg-approxSSSP}.
	\hfill $\Box$
	\medskip
	
	\noindent
	The correctness of Algorithm~\ref{alg-approxSSSP} is thus proved.
	Later, the first statement of Lemma~\ref{lemma-consistency} will also be used to obtain an efficient implementation of the approximate update.
	
	Next, we consider the time complexity of Algorithm~\ref{alg-approxSSSP}.
	Using the same argument as in Section~\ref{sec-exact}, we see that the running time of Algorithm~\ref{alg-approxSSSP} without line~10-12 is $O(n \log n)$.
	Line~10 can be implemented using the same method as in Section~\ref{sec-correction}, namely building a WVD on the points in $A_{\boxplus_c} \backslash A_{\Box_c}$ and querying for each point in $A_{\Box_c}$ (the correctness follows from the argument in Section~\ref{sec-correction}).
	Also, line~11 can be implemented in this way, because the points in $A_{\Box_c}$ are pairwise adjacent in $G$.
	Therefore, the total running time for line~10 and 11 is $O(n \log n)$.
	It suffices to analyze the time cost of line~12, the approximate update.
	
	\subsection{Approximate Update} \label{sec-update'}
	In order to implement the approximate update (line~12) in Algorithm~\ref{alg-approxSSSP}, we (implicitly) build another grid $\varGamma'$ on the plane, which consists of square cells with side-length $\varepsilon/8$.
	To avoid confusion, we use $\blacksquare$ to denote a cell in $\varGamma'$.
	For a point $a \in S$, let $\blacksquare_a$ denote the cell in $\varGamma'$ containing $a$.
	For a set $P$ of points in $\mathbb{R}^2$ and a cell $\blacksquare$ in $\varGamma'$, define $P_\blacksquare = P \cap \blacksquare$.
	
	Line~12 of Algorithm~\ref{alg-approxSSSP} is \textsc{ApproxUpdate}$(A_{\Box_c},A_{\boxplus_c} \backslash A_{\Box_c})$.
	Let $U = A_{\Box_c}$ and $V = A_{\boxplus_c} \backslash A_{\Box_c}$.
	We shall use two special properties of the set $U$: \textbf{(i)} all the points in $U$ are contained in one cell in $\varGamma$ and \textbf{(ii)} $\text{dist}[u] \leq \text{dist}[u'] + \lVert u'-u \rVert$ for all $u,u' \in U$ before the procedure \textsc{ApproxUpdate}$(U,V)$, which follows from Lemma~\ref{lemma-consistency}.
	Our algorithm for implementing \textsc{ApproxUpdate}$(U,V)$ is shown in Algorithm~\ref{alg-approxupdate}, which is a variant of Algorithm~\ref{alg-refine}.
	Here we no longer need the $\text{dist}'[\cdot]$ table because $U$ and $V$ are disjoint.
	
	\begin{algorithm}[tbph]
		\caption{\textsc{ApproxUpdate}$(U,V)$}
		\begin{algorithmic}[1]
			%\State $\text{dist}'[u] \leftarrow \text{dist}[u]$ for all $u \in U$.
			\State Sort the points in $U = \{u_1,\dots,u_{|U|}\}$ such that $\text{dist}[u_1] \leq \cdots \leq \text{dist}[u_{|U|}]$
			\For{$i=1,\dots,|U|$}
			\State $V_i \leftarrow \{v \in V: v \in \odot_{u_i} \text{ and } v \notin \odot_{u_j} \text{ for all } j<i\}$
			\EndFor
			\State $U' \leftarrow \{u_j : j \geq k \text{ for all } k \text{ such that } u_k \in \blacksquare_{u_j}\}$
			\State $B \leftarrow \emptyset$        
			\For{$i=|U|,\dots,1$}
			\IIf{$u_i \in U'$} $B \leftarrow B \cup \{u_i\}$
			%\State $B \leftarrow B \cup \{u_i\}$
			%\EndIf
			\For{$v \in V_i$}
			\State $p \leftarrow \arg \min_{b \in B} \{\text{dist}[b] + \lVert b-v \rVert\}$
			\IIf{$p \notin \odot_v$} $p \leftarrow u_i$
			\If{$\text{dist}[v] > \text{dist}[p] + \lVert p-v \rVert$}
			\State $\text{dist}[v] \leftarrow \text{dist}[p] + \lVert p-v \rVert$
			\State $\text{pred}[v] \leftarrow p$
			%\nolinenumbers
			\EndIf
			\EndFor
			\EndFor
		\end{algorithmic}
		\label{alg-approxupdate}
	\end{algorithm}
	
	Recall the definition of the sub-routine \textsc{ApproxUpdate} in Section~\ref{sec-approx}.
	To verify the correctness of Algorithm~\ref{alg-approxupdate}, it suffices to show that just before line~11, the point $p$ satisfies that $p \in U \cap \odot_v$ and $\text{dist}[p] + \lVert p-v \rVert \leq \text{dist}[r] + \lVert r-v \rVert + \varepsilon/2$ for all $r \in U \cap \odot_v$.
	The condition $p \in \odot_v$ is clearly satisfied, because of line~10 (note that $u_i \in \odot_v)$.
	To verify the latter condition, we first observe the following fact, which is directly implied by the property \textbf{(ii)} of $U$ mentioned in the beginning of Section~\ref{sec-update'}.
	
	\begin{fact} \label{fact-distance}
		For all $a \in \mathbb{R}^2$ and $u,u' \in U$, we have, just before line~11 of Algorithm~\ref{alg-approxupdate},
		\begin{equation*}
		|(\textnormal{dist}[u] + \lVert u-a \rVert) - (\textnormal{dist}[u'] + \lVert u'-a \rVert)| \leq 2 \lVert u-u' \rVert.
		\end{equation*}
	\end{fact}
	\textit{Proof.}
	By the property \textbf{(ii)} of $U$, we have $\text{dist}[u]-\text{dist}[u'] \leq \lVert u-u' \rVert$.
	Thus,
	\begin{equation*}
	(\text{dist}[u] + \lVert u-a \rVert) - (\text{dist}[u'] + \lVert u'-a \rVert) \leq \lVert u-u' \rVert + \lVert u-a \rVert - \lVert u'-a \rVert \leq 2 \lVert u-u' \rVert,
	\end{equation*}
	where the second ``$\leq$'' follows from the triangle inequality.
	Symmetrically, we can also show that $(\text{dist}[u'] + \lVert u'-a \rVert) - (\text{dist}[u] + \lVert u-a \rVert) \leq 2 \lVert u-u' \rVert$.
	\hfill $\Box$
	\medskip
	
	%\textit{Proof.}
	%For a cell $\blacksquare$ in $\varGamma'$, define $\rho(\blacksquare) = \max\{j: u_j \in \blacksquare\}$.
	%Let $U' = \{u_j: j = \rho(\blacksquare)\}$.
	%In other words, $U'$ is the subset of $U$ consisting of the points who have the largest indices among all the points belonging to the same $\varGamma'$-cell.
	%One can easily verify that $B = \{u_i,\dots,u_{|U|}\} \cap U'$ (just before line~8).
	%Therefore, if $j \geq i$, we have $u_k \in B$ for $k = \rho(\blacksquare_{u_j})$.
	
\begin{figure}[t]
		\begin{minipage}[t]{\linewidth}
			\begin{center}
				\includegraphics[totalheight=1.5in]{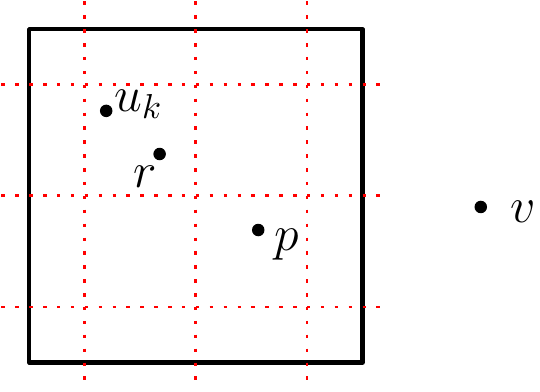}
				\caption{\footnotesize
					Illustrating some points in the proof of Lemma~\ref{lem:approx}. The solid square is $\Box_c$ and the dotted grid is $\varGamma'$. The two points $u_k$ and $r$ (i.e., $u_j$) are in the same cell of $\varGamma'$.}
				\label{fig:grid}
			\end{center}
		\end{minipage}
	\end{figure}

	\noindent
	With the above fact, we can prove the following lemma.
	\begin{lemma}\label{lem:approx}
		Just before line~11 of Algorithm~\ref{alg-approxupdate}, we have $\textnormal{dist}[p] + \lVert p-v \rVert \leq \textnormal{dist}[r] + \lVert r-v \rVert + \varepsilon/2$ for all $r \in U \cap \odot_v$.
	\end{lemma}
	\textit{Proof.}
	Suppose we are at the moment just before line~11 of Algorithm~\ref{alg-approxupdate}.
	We want to prove $\text{dist}[p]+\lVert p-v \rVert \leq \text{dist}[r]+\lVert r-v \rVert$ for all $r \in U \cap \odot_v$.
	For convenience, we write $h(u) = \textnormal{dist}[u] + \lVert u-v \rVert$ for all $u \in U$.
	Fact~\ref{fact-distance} implies that $|h(u)-h(u')| \leq 2 \lVert u-u' \rVert$ for all $u,u' \in U$.
	Consider a point $r \in U \cap \odot_v$ and assume $r = u_j$.
	It suffices to show $h(p) \leq h(r) + \varepsilon/2$.
	Note that $j \geq i$ since $u_1,\dots,u_{i-1} \notin \odot_v$.
	Set $k = \max\{t: u_t \in \blacksquare_{u_j}\}$, e.g., see Fig.~\ref{fig:grid}.
	We have $k \geq j \geq i$ and $u_k \in U'$, which implies $u_k \in B$.
	%Let $\tilde{p}$ be the point $p$ just after line~10.
	%We first show that $h(\tilde{p}) \leq h(r) + \varepsilon/2$.
	To prove $h(p) \leq h(r) + \varepsilon/2$, we distinguish two cases: $p$ is not changed in line~10 and $p$ is changed in line~10.

	\begin{itemize}
		\item
		Assume $p$ is not changed in line~10.
		Then $p = \arg\min_{b \in B} h(b)$ and in particular $h(p) \leq h(u_k)$.
		Therefore, 
		\begin{equation*}
		h(p)-h(r) \leq h(u_k)-h(r) \leq |h(u_k)-h(r)| \leq 2 \lVert u_k-r \rVert \leq \varepsilon/2,    
		\end{equation*}
		where the last inequality holds because $u_k\in \blacksquare_r$ and the side-length of $\blacksquare_r$ is $\varepsilon/8$.
		
		\item
		Assume $p$ is changed in line~10.
		Then $p = u_i$.
		Let $\tilde{p} = \arg\min_{b \in B} h(b)$.
		Using the same argument as above, we can deduce $h(\tilde{p}) \leq h(r) + \varepsilon/2$.
		Also, we have $\tilde{p} \notin \odot_v$ and hence $\lVert \tilde{p}-v \rVert >1 \geq \lVert p-v \rVert$, for otherwise $p$ would not be changed in line~10.
		Furthermore, $\text{dist}[p] = \text{dist}[u_i] \leq \text{dist}[\tilde{p}]$ because $\tilde{p} \in B$ and $B \subseteq \{u_i,\dots,u_{|U|}\}$.
		As such,
		\begin{equation*}
		h(p) = \text{dist}[p] + \lVert p-v \rVert \leq \text{dist}[\tilde{p}] + \lVert \tilde{p}-v \rVert = h(\tilde{p}) \leq h(r) + \varepsilon/2.
		\end{equation*}
	\end{itemize}
	
	This completes the proof of the lemma.
	\hfill $\Box$
	\medskip
	
	To see the time complexity of Algorithm~\ref{alg-approxupdate}, let $k = |U|+|V|$.
	The sorting in line~1 takes $O(|U| \cdot \log |U|)$ time.
	The loop in line~2-3 can be implemented in $O(k \log |U|)$ time using the same method as in Section~\ref{sec-refine}.
	In line~4, we can compute the set $U'$ in $O(|U| \cdot \log (|S_{\boxplus_c}|/\varepsilon))$ time by grouping the points in $U$ that belong to the same $\varGamma'$-cell (see Appendix~\ref{sec:locategrid} for a more detailed discussion).
	The loop in line~6-13 is basically weighted nearest-neighbor search (line~9) with insertions (line~7).
	There are $O(|V|)$ queries and $O(|U'|)$ insertions.
	Note that $|U'| = O(\varepsilon^{-2})$, because of the property \textbf{(i)} of $U$.
	Therefore, if we use $f(k_1,k_2)$ to denote the time cost for solving the OIWNN problem with $k_1$ operations in which at most $k_2$ operations are insertions, then the loop in line~6-13 takes $f(k,O(\varepsilon^{-2}))$ time.
	In sum, the running time of Algorithm~\ref{alg-approxupdate} is $O(f(k,O(\varepsilon^{-2})) + k \log k)$ time.
	Therefore, the approximate update in Algorithm~\ref{alg-approxSSSP} can be done in $O(f(|S_{\boxplus_c}|,O(\varepsilon^{-2}))+|S_{\boxplus_c}| \cdot \log (|S_{\boxplus_c}|/\varepsilon))$ time.
	
	\subsection{Putting Everything Together}
	Except the approximate update, Algorithm~\ref{alg-SSSP} runs in $O(n \log n)$ time.
	Section~\ref{sec-update'} shows that the approximate update of each iteration can be done in $O(f(k,O(\varepsilon^{-2})) + k \log k+k\log(1/\varepsilon))$ time where $k = |A_{\Box_c}| + |A_{\boxplus_c}| = O(|S_{\boxplus_c}|)$ and $f(k_1,k_2)$ is the time for solving the OIWNN problem with $k_1$ operations in which at most $k_2$ operations are insertions.
	Noting the fact $\sum_{i=1}^m |S_{\boxplus_{c_i}}| \leq 25 n$, we can conclude the following.
	\begin{theorem}\label{thm-approx}
		Suppose the OIWNN problem with $k_1$ operations in which at most $k_2$ operations are insertions can be solved in $f(k_1,k_2)$ time, and assume $f(k_1,k_2)/k_1$ is a non-decreasing function of $k_1$ for any fixed $k_2$.
		Then there exists a $(1+\varepsilon)$-approximate SSSP algorithm in weighted unit-disk graphs with $O(n \log n + n \log (1/\varepsilon) + f(n,O(\varepsilon^{-2})))$ running time, where $n$ is the number of the vertices.
	\end{theorem}
	\textit{Proof.}
	According to our analysis and the fact $\sum_{i=1}^m |S_{\boxplus_{c_i}}| \leq 25 n$, the overall running time of Algorithm~\ref{alg-approxSSSP} is $O(n \log n + n \log (1/\varepsilon) + \sum_{i=1}^m f(|S_{\boxplus_{c_i}}|,O(\varepsilon^{-2})))$.
	Since $f(k_1,k_2)/k_1$ is a non-decreasing function of $k_1$ (for a fixed $k_2$), we have
	\begin{equation*}
	\sum_{i=1}^m f(|S_{\boxplus_{c_i}}|,O(\varepsilon^{-2})) = \sum_{i=1}^m |S_{\boxplus_{c_i}}| \cdot \frac{f(|S_{\boxplus_{c_i}}|,O(\varepsilon^{-2}))}{|S_{\boxplus_{c_i}}|} \leq \sum_{i=1}^m |S_{\boxplus_{c_i}}| \cdot \frac{f(n,O(\varepsilon^{-2}))}{n} \leq 25 f(n,O(\varepsilon^{-2})).
	\end{equation*}
	Therefore, Algorithm~\ref{alg-approxSSSP} runs in $O(n \log n + n \log (1/\varepsilon) + f(n,O(\varepsilon^{-2})))$ time.
	\hfill $\Box$
	\medskip
	
	\noindent
	We give in Appendix~\ref{appx-WNN} a linear-space algorithm with $f(k_1,k_2) = O(k_1 \log^2 k_2)$.
	%that solves in $O(k_1 \log^2 k_2)$ time the OIWNN problem with $k_1$ operations in which at most $k_2$ operations are insertions, which implies $f(k_1,k_2) = O(k_1 \log^2 k_2)$.
	By plugging in this algorithm, we can obtain the following corollary.
	%implement Algorithm~\ref{alg-approxSSSP} in $O(n \log n + n \log^2 (1/\varepsilon))$ time and linear space, which results in 
	\begin{corollary} \label{cor-approximation}
		For any $\varepsilon>0$, there exists a $(1+\varepsilon)$-approximate SSSP algorithm in weighted unit-disk graphs with $O(n \log n + n \log^2 (1/\varepsilon))$ time and $O(n)$ space, where $n$ is the number of the vertices.
	\end{corollary}
	
	\subsection{Improved Distance Oracles in Weight Unit-Disk Graphs}
	As an application of our $(1+\varepsilon)$-approximate SSSP algorithm in weighted-unit disk graphs presented in Section~\ref{sec-approx}, we improve the preprocessing time of the additively-approximate and $(1+\varepsilon)$-approximate distance oracles for weighted unit-disk graphs given by Chan and Skrepetos~\cite{chan2018approximate}.
	
	Let $G = (V,E)$ be a weighted unit-disk graph with $n$ vertices, and $\Delta = \max_{s,t \in V} d_G(s,t)$ be its diameter.
	It was shown in \cite{chan2018approximate} that, for any $\varepsilon>0$ such that $\varepsilon \Delta \geq 1$, if the $(1+\varepsilon)$-approximate SSSP problem in $G$ can be solved in $T(n)$ time, then one can construct in $O((1/\varepsilon) T(n) \log n)$ time an additively-approximate distance oracle for $G$ with stretch $O(\varepsilon\Delta)$; the distance oracle uses $O((1/\varepsilon) n \log n)$ space and $O((1/\varepsilon) \log n)$ query time.
	By plugging in our algorithm in Corollary~\ref{cor-approximation}, we conclude the following.
	\begin{theorem} \label{thm-additive}
		Given a weighted unit-disk graph $G$ with diameter $\Delta$, for any $\varepsilon>0$ such that $\varepsilon \Delta \geq 1$, one can build in $O((1/\varepsilon) n \log^2 n + (1/\varepsilon) n \log n \log^2(1/\varepsilon))$ time an $O(\varepsilon\Delta)$-stretch additively-approximate distance oracle for $G$ with $O((1/\varepsilon) n \log n)$ space and $O((1/\varepsilon) \log n)$ query time, where $n$ is the number of the vertices of $G$.
	\end{theorem}
	The above theorem improves the $O((1/\varepsilon)^3 n \log^2 n)$ preprocessing time of the additive-approximate distance oracle given by Chan and Skrepetos~\cite{chan2018approximate}.
	
	Chan and Skrepetos~\cite{chan2018approximate} further showed that an $O(\varepsilon\Delta)$-stretch additively-approximate distance oracle $\mathcal{D}_\text{add}$ described in Theorem~\ref{thm-additive} can be used to build a $(1+\varepsilon)$-approximate distance oracle $\mathcal{D}_\text{approx}$ for weighted unit-disk graphs with $O((1/\varepsilon) n \log^2 n + (1/\varepsilon)^4 n)$ space and $O((1/\varepsilon) \log^2 n)$ query time.
	The preprocessing time of $\mathcal{D}_\text{approx}$ is $O(T'(n) \log n + (1/\varepsilon)^6 n \log (1/\varepsilon))$ where $T'(n)$ is the time for building $\mathcal{D}_\text{add}$.
	By Theorem~\ref{thm-additive}, we have $T'(n) = O((1/\varepsilon) n \log^2 n + (1/\varepsilon) n \log n \log^2(1/\varepsilon))$.
	Thus, we conclude the following.
	\begin{theorem}
		Given a weighted unit disk graph $G$, for any $\varepsilon>0$, one can build in $O((1/\varepsilon) n \log^3 n + (1/\varepsilon) n \log^2 n \log^2(1/\varepsilon) + (1/\varepsilon)^6 n \log (1/\varepsilon))$ time a $(1+\varepsilon)$-approximate distance oracle for $G$ with $O((1/\varepsilon) n \log^2 n + (1/\varepsilon)^4 n)$ space and $O((1/\varepsilon) \log^2 n)$ query time, where $n$ is the number of the vertices of $G$.
	\end{theorem}
	The above theorem improves the $O((1/\varepsilon)^3 n \log^3 n + (1/\varepsilon)^6 n \log (1/\varepsilon))$ preprocessing time of the $(1+\varepsilon)$-approximate distance oracle given by Chan and Skrepetos~\cite{chan2018approximate}.
	
	\medskip
	\noindent
	\textbf{Acknowledgement.}
	The authors would like to thank Timothy Chan for the discussion, and in particular, for suggesting the algorithm for Lemma~\ref{lem:firstLoopAlgo1}.
	Jie Xue would like to thank his advisor Ravi Janardan for his consistent advice and support.
	
	\bibliography{main.bib}

\begin{thebibliography}{10}

\bibitem{ref:BentleyDe79}
Jou~L. Bentley.
\newblock Decomposable searching problems.
\newblock {\em Information Processing Letters}, 8:244--251, 1979.

\bibitem{cabello2015shortest}
Sergio Cabello and Miha Jej{\v{c}}i{\v{c}}.
\newblock Shortest paths in intersection graphs of unit disks.
\newblock {\em Computational Geometry: Theory and Applications}, 48:360--367,
  2015.

\bibitem{chan2016all}
Timothy~M Chan and Dimitrios Skrepetos.
\newblock All-pairs shortest paths in unit-disk graphs in slightly subquadratic
  time.
\newblock In {\em Proceedings of the 27th International Symposium on Algorithms
  and Computation (ISAAC)}, pages 24:1--24:13, 2016.

\bibitem{chan2017all}
Timothy~M. Chan and Dimitrios Skrepetos.
\newblock All-pairs shortest paths in geometric intersection graphs.
\newblock In {\em Proceedings of the 15th Algorithms and Data Structures
  Symposium (WADS)}, pages 253--264, 2017.

\bibitem{chan2018approximate}
Timothy~M Chan and Dimitrios Skrepetos.
\newblock Approximate shortest paths and distance oracles in weighted unit-disk
  graphs.
\newblock In {\em Proceedings of the 34th International Symposium on
  Computational Geometry (SoCG)}, pages 24:1--24:13, 2018.

\bibitem{clark1990unit}
Brent~N. Clark, Charles~J. Colbourn, and David~S. Johnson.
\newblock Unit disk graphs.
\newblock {\em Discrete Mathematics}, 86:165--177, 1990.

\bibitem{deBerg2016fine}
Mark de~Berg, Kevin Buchin, Bart~M.P. Jansen, and Gerhard Woeginger.
\newblock Fine-grained complexity analysis of two classic {TSP} variants.
\newblock In {\em Proceedings of the 43rd International Colloquium on Automata,
  Languages, and Programming (ICALP)}, pages 5:1--5:14, 2016.

\bibitem{EdelsbrunnerOp86}
Herbert Edelsbrunner, Leonidas~J. Guibas, and Jorge Stolfi.
\newblock Optimal point location in a monotone subdivision.
\newblock {\em SIAM Journal on Computing}, 15:317--340, 1986.

\bibitem{ref:FortuneA87}
Steven Fortune.
\newblock A sweepline algorithm for {Voronoi} diagrams.
\newblock {\em Algorithmica}, 2:153--174, 1987.

\bibitem{gao2005well}
Jie Gao and Li~Zhang.
\newblock Well-separated pair decomposition for the unit-disk graph metric and
  its applications.
\newblock {\em SIAM Journal on Computing}, 35:151--169, 2005.

\bibitem{kaplan2017dynamic}
Haim Kaplan, Wolfgang Mulzer, Liam Roditty, Paul Seiferth, and Micha Sharir.
\newblock Dynamic planar voronoi diagrams for general distance functions and
  their algorithmic applications.
\newblock In {\em Proceedings of the 28th Annual ACM-SIAM Symposium on Discrete
  Algorithms (SODA)}, pages 2495--2504, 2017.

\bibitem{matsui1998approximation}
Tomomi Matsui.
\newblock Approximation algorithms for maximum independent set problems and
  fractional coloring problems on unit disk graphs.
\newblock In {\em Proceedings of Japanese Conference on Discrete and
  Computational Geometry (JCDCG)}, pages 194--200, 1998.

\bibitem{roditty2011bounded}
Liam Roditty and Michael Segal.
\newblock On bounded leg shortest paths problems.
\newblock {\em Algorithmica}, 59:583--600, 2011.

\end{thebibliography}
	
	\newpage
	\appendix
	
	%\bigskip
	\noindent
	{\LARGE \bf Appendix}
	
	\section{Locating Points in a Grid}
	\label{sec:locategrid}
	
	Let $\varGamma$ be a grid on the plane consisting of square cells with side-length $\gamma$.
	Assume the origin is a grid point of $\varGamma$.
	Given a point $a \in \mathbb{R}^2$, the cell in $\varGamma$ containing $a$, $\Box_a$, can be computed directly using the $\mathsf{floor}$ function, because $\Box_a = [\lfloor x_a/\gamma \rfloor,\lfloor x_a/\gamma \rfloor + \gamma] \times [\lfloor y_a/\gamma \rfloor,\lfloor y_a/\gamma \rfloor + \gamma]$ where $(x_a,y_a)$ is the coordinate of $a$.
	However, many fundamental computational models do not assume the existence of a constant-time $\mathsf{floor}$ function.
	We show that, at least in our algorithms, we can locate points in a grid using only basic arithmetic operations without influencing their performances.
	
	We first notice that the $\mathsf{floor}$ function can be simulated using basic arithmetic operations.
	\begin{fact}
		Given a real number $r \geq 0$, one can compute $\lfloor r \rfloor$ in $O(\log r + 1)$ time using only basic arithmetic operations on real numbers.
	\end{fact}
	\textit{Proof.}
	We first compute the smallest integer $k$ such that $r < 2^k$ in $O(\log r + 1)$ time.
	Note that $k = O(\log r+1)$.
	Let $U = 2^k$.
	If $U = 1$, then $\lfloor r \rfloor = 0$.
	Otherwise, we set $r_0 = 0$ and repeat the following procedure until $U = 1$.
	\begin{enumerate}
		\item $U \leftarrow U/2$.
		\item If $r \geq U$, then $r \leftarrow r-U$ and $r_0 \leftarrow r_0+U$.
	\end{enumerate}
	It is clear that eventually $r_0 = \lfloor r \rfloor$ and the above procedure takes $O(k)$ time.
	\hfill $\Box$
	\medskip
	
	\noindent
	Therefore, for a point $a \in [0,U] \times [0,U]$, one can compute $\Box_a$ in $O(\log (U/\gamma)+1)$ time.
	
	In our SSSP algorithms, we can in fact assume that all the points in $S$ lie in the square $K = [0,2n] \times [0,2n]$.
	Indeed, via a translation, we can make $s = (n,n)$.
	Then any point $a \notin K$ is not contained in the connected components of $G$ containing $s$, i.e., $d_G(s,a) = \infty$.
	Indeed, if $a \notin K$, then $n-1 < \lVert s-a \rVert \leq d_G(s,a)$, which implies $d_G(s,a) = \infty$ because the length of any simple path in a weighted unit-disk graph is at most $n-1$.
	With the assumption that $S \subseteq K$, we can locate each point in a grid in $O(\log n + \log(1/\gamma))$ time where $\gamma$ is the side-length of a grid cell.
	There are two grids $\varGamma$ and $\varGamma'$ used in our algorithms with $\gamma = 1/2$ and $\gamma = \varepsilon/8$, respectively.
	Thus, locating all the $n$ points in $\varGamma$ (resp., $\varGamma'$) takes $O(n \log n)$ (resp., $O(n \log n + n \log(1/\varepsilon))$) time.
	
	\section{Offline Insertion-Only Weighted Nearest-Neighbor Problem} 
	\label{appx-WNN}
	
	In this section, we show that the (2D) offline insertion-only (additively-)weighted nearest-neighbor (OIWNN) problem with $n$ operations (i.e., insertions and queries) in which at most $m$ operations are insertions can be solved in $O(n \log^2 m)$ time using linear space.
	In particular, the OIWNN problem with $n$ operations can be solved in $O(n \log^2 n)$ time using linear space.
	
	Suppose we are given a sequence $(o_1,\dots,o_n)$ of $n$ operations consisting of $m$ insertions and $n-m$ queries.
	Let $a_i$ denote the weighted point in $\mathbb{R}^2$ inserted by the $i$-th insertion for $i \in \{1,\dots,m\}$.
	For each query $Q$ in the sequence $(o_1,\dots,o_n)$, we denote by $\mathsf{ans}(Q)$ the true answer of $Q$, which is a point in $\{a_1,\dots,a_m\}$.
	To solve the problem, we split the sequence $(o_1,\dots,o_n)$ into two (consecutive) sub-sequences $(o_1,\dots,o_k)$ and $(o_{k+1},\dots,o_n)$ each of which consists of $m/2$ insertions.
	We regard each sub-sequence as a sub-problem and solve it recursively.
	After this, we build a (additively-)weighted Voronoi Diagram (WVD) on $\{a_1,\dots,a_{m/2}\}$, which takes $O(m \log m)$ time.
	Consider a query $Q$ among $o_1,\dots,o_k$.
	By solving the sub-problem $(o_1,\dots,o_k)$, we obtain an answer for $Q$, which is the same as its answer in the original problem, i.e., $\mathsf{ans}(Q)$.
	Consider a query $Q$ among $o_{k+1},\dots,o_n$.
	Let $q \in \mathbb{R}^2$ be the corresponding query point.
	Suppose $Q$ is in between the $i$-th insertion and the $(i+1)$-th insertion, where $i \geq m/2$.
	By solving the sub-problem $(o_{k+1},\dots,o_n)$, we obtain an answer $a$ for $Q$, which is the weighted nearest neighbor of $q$ in $\{a_{m/2+1},\dots,a_i\}$.
	We then further query the WVD to obtain the weighted nearest neighbor $a'$ of $q$ in $\{a_1,\dots,a_{m/2}\}$.
	With $a$ and $a'$ in hand, we can directly compute the weighted nearest neighbor of $q$ in $\{a_1,\dots,a_i\}$, which is just $\mathsf{ans}(Q)$.
	In this way, we obtain the answers for all the $n-m$ queries and solve the problem.
	
	We now analyze the running time of the above algorithm.
	Except the time for recursively solving the two sub-problems, the algorithm uses $O(n \log m)$ time, because building the WVD takes $O(m \log m)$ time and querying the WVD takes $O((n-m) \log m)$ time in total.
	Let $T(n,m)$ be the time cost for handling a sequence of $n$ operations in which $m$ operations are insertions.
	We then have the recurrence
	\begin{equation*}
	T(n,m) = T(k,m/2) + T(n-k,m/2) + O(n \log m).
	\end{equation*}
	The depth of the recurrence is $O(\log m)$, and the time cost in each level is $O(n \log m)$.
	Therefore, we have $T(n,m) = O(n \log^2 m)$.
	Now we see that our algorithm solves in $O(n \log^2 m)$ time the OIWNN problem with $n$ operations in which \textit{exactly} $m$ operations are insertions.
	This further implies that the OIWNN problem with $n$ operations in which at most $m$ operations are insertions can be solved in $O(n \log^2 m)$ time.
	Finally, it is easy to see that our algorithm above only requires linear space (with a careful implementation).
	
\end{document}